\documentclass[5p,twocolumn,10pt]{elsarticle}

\usepackage{amsmath}
\usepackage{hyperref}
\usepackage{soul}
\usepackage{lineno}

\usepackage{multirow}
\usepackage{makecell}
\usepackage{epstopdf}
\usepackage{amsmath}%
\usepackage{amsfonts}%
\usepackage{amssymb}%
\usepackage{graphicx}
\usepackage{mathrsfs}
\usepackage{algorithm}
\usepackage[noend]{algpseudocode}
\usepackage{tikz-cd}
\usepackage{url}
\usepackage{arydshln}
\usepackage{fancyhdr}
\usepackage{array}
\usepackage{lineno}

\modulolinenumbers[5]

\makeatletter
\def\BState{\State\hskip-\ALG@thistlm}
\makeatother

\usepackage{comment} 
\usepackage{subfig}
\usepackage{hyperref}
\hypersetup{colorlinks=true,allcolors=blue}
\usepackage{hypcap}

\newcolumntype{C}{>{\centering\arraybackslash} m	 } 

\newcommand{\svnidlong}[4]{}%
\newcommand{\ch}[1]{{\color{black}{#1}}}
\newcommand{\x}[1]{{\color{black}{#1}}}
\newcommand{\q}[1]{{\color{black}{#1}}}

\begin{document}
\baselineskip 11pt

\begin{frontmatter}

\title{Scalable Path Level Thermal History Simulation of PBF process validated by Melt Pool Images}

\author[1]{Xin Liu}
\ead{xin@intact-solutions.com}

\author[1]{Xingchen Liu\corref{cor1}}
\ead{xliu@intact-solutions.com}

\author[1]{Goldy Kumar}
\ead{gkumar@intact-solutions.com}

\author[2]{Paul Witherell}
\ead{paul.witherell@nist.gov}

\cortext[cor1]{Corresponding author}
\address[1]{Intact Solutions, Inc.}
\address[2]{National Institute of Standards and Technology}

\begin{abstract}

\ch{The thermal history of the powder bed fusion (PBF) process is critical due to its significant influences on the material properties, residual stress, and part warping or distortion. These quantities heavily affect the quality and performance of parts. } The contact-aware path-level (CAPL) discretization approach~\cite{zhang2018linear,zhang2019towards,zhang2022scalable} was recently proposed to support scalable thermal simulation at the path level of AM processes driven by a moving heat source. Compared to other thermal simulation approaches, CAPL tailors discretization to the manufacturing toolpath and adopts locality for linear time complexity in part-scale thermal history simulations. This approach essentially uses scalable simulations to simulate the fabrication process of a part through the aggregation of melt pools, scan paths\q{, and layers}. 

In this paper we outline the development of a scalable PBF thermal history simulation built on CAPL and based on melt pool physics and dynamics. The new approach inherits linear scalability from CAPL and has three novel ingredients. Firstly, to simulate the laser scanning on a solid surface, we discretize the entire simulation domain instead of only the manufacturing toolpath by appending fictitious paths to the manufacturing toolpath. Secondly, to simulate the scanning on overlapping toolpaths, the path-scale simulations are initialized by a Voronoi diagram for line segments discretized from the manufacturing toolpath. Lastly, we propose a modified conduction model that considers the high thermal gradient around the melt pool. We validate the simulation against melt pool images captured with the co-axial melt pool monitoring (MPM) system on the NIST Additive Manufacturing Metrology Testbed (AMMT). Excellent agreements in the length and width of melt pools are found between simulations and experiments conducted on a custom-controlled laser powder bed fusion (LPBF) testbed on a nickel-alloy (IN625) solid surface. To the authors’ best knowledge, this paper is the first to validate a full path-scale thermal history with experimentally acquired melt pool images. Comparing the simulation results and the experimental data, we discuss the influence of laser power on the melt pool length on the path-scale level. We also identify the possible ways to further improve the accuracy of the CAPL simulation without sacrificing efficiency.

\end{abstract}


\end{frontmatter}

\section{Introduction}

Laser powder bed fusion (LPBF) additive manufacturing (AM) uses lasers as moving heat sources to melt and solidify thin layers of metal powder along a predefined tool path layer by layer \cite{vock2019powders, bhavar2017review, singh2021powder}.  \ch{Many complex multi-physics and multi-phase phenomena are involved in the different stages of the LPBF process, including the initial melting of metal powders, the fluid dynamics and heat transfer of the molten metal, and the stresses and microscopic grain structure resulting from the solidification of liquid metals. These phenomena have important influences on the quality and performance of the manufactured part. As many of them are related to the thermal history of the part, an accurate and efficient thermal history simulation becomes critical to the understanding and improvement of LPBF processes.
For example, thermal history influences the microstructure ~\cite{acharya2017prediction} and the material properties ~\cite{hilaire2019high,inaekyan2019microstructure}.} The thermal history also influences the part’s geometric accuracy through the residual stress, which is caused by uneven thermal expansion and shrinkage~\cite{luo2018survey}, which is closely related to the thermal history of LPBF parts. 

The thermal history of a specific part is the result of its manufacturing process plan. The process plan includes processing parameters (e.g., the environmental and preheating temperature of the platform) and path-dependent scanning information including the power and speed of the laser. For a given geometric model, the build direction is determined first. Based on the process specifications such as the powder layer thickness, the geometric model is then sliced into layers normal to the building direction. Finally, the layers from slicing are filled with scanning paths, which direct the laser to scan the path segments with the given laser power. The scanning paths have a critical influence on thermal history. For example, a given location might be melted once or remelted multiple times if different scanning paths are adopted. For a given geometric model, different combinations of possible building directions and scanning path results in an enormous degree of freedom on the path-scale level for process plan designing. The complexity of scanning patterns and the utilization of fast-moving and high-energy lasers in LPBF processes lead to complex heating/cooling and phase transitions, which complicates the thermal history across the entire part. Path-scale simulation approaches are critical to predicting the thermal history and exploring different process plans.

We develop a scalable PBF thermal simulation approach based on CAPL. CAPL tailors discretization to the manufacturing toolpath and adopts locality for linear time complexity in part-scale thermal history simulations. The new approach \q{(will be referred as PBF-CAPL)} inherits linear scalability from CAPL and has three novel ingredients. Firstly, to simulate the laser scanning on a solid surface, we discretize the entire simulation domain instead of only the manufacturing toolpath by appending the fictitious paths to the manufacturing toolpath. Secondly, to simulate the scanning on the overlapping toolpath, the element widths are initialized by a Voronoi diagram of the manufacturing toolpath. Lastly, we propose a modified conduction model that considers the high thermal gradient around the melt pool. We validate the new approach against melt pool images captured with the co-axial melt pool monitoring (MPM) system on the Manufacturing Metrology Testbed (AMMT) developed at the National Institute of Standards and Technology (NIST)~\cite{lane2016design, yeung2020residual}. Excellent agreements in the length and width of melt pools are found between simulations and experiments conducted on a custom-controlled laser powder bed fusion (LPBF) testbed on a nickel-alloy (IN625) solid surface.

The rest of the paper is organized as follows. \ch{We review the related work in the next section.} For the sake of completeness, we provide an overview of the original CAPL approach in Section \ref{sec:CAPL}. In Section \ref{sec:CAPLlimit}, \q{we identify and implement the needed improvements to account for \ch{the limitations of the original CAPL} and specifics of the validation dataset. }

In Section \ref{sec:validation}, we validate the simulation results against the experimental dataset that includes scan paths with varying laser power, where we find good agreement in the melt pool shapes between simulation results and the experimental dataset. In addition to comparing the melt pool images frame by frame, we also analyze various trends observed in the dataset. These include the influence of laser power on melt pool length, the melt pool length evolution on the same scanning vector, and scan-wise melt pool length evolution. 

Based on the observation in Section \ref{sec:validation}, Section \ref{sec:conclusion} discusses \ch{further} improvements to CAPL-based thermal history simulation. These include the incorporation of the laser absorptivity model affected by surface roughness and keyholes and utilizing machine learning to determine melt pool shapes. 

\section{Related Work}

\subsection{LPBF in-process temperature measurements}

Two main types of experimental temperature measurements in the LPBF process are often used for in-process temperature measurements in LPBF: (a) Thermocouples and devices which can measure temperature at specific locations ~\cite{denlinger2017thermomechanical, schnell2021experimental}. (b) ``co-axial'' approaches \cite{goossens2021virtual} or ``off-axis'' approaches \cite{mitchell2020linking} thermal camera imaging and pyrometry devices that can measure heatmap and heating/cooling profiles \cite{lu2020camera} of the layer or the local region around the melt pool ~\cite{grasso2021situ}. \q{Such direct temperature measurement of melt pool temperature measurement usually involves the estimation of melt pool radiation, which can be modeled by the Sakuma-Hattori equation \cite{saunders2003interpolation}. In such measurements, it is required to have the delicate calibration of surface properties such as emissivity, which is difficult to estimate due to fast phase transitions like evaporation or liquidation ~\cite{liebmann2021exploration}. In the LPBF process, researchers tried to directly use the raw data. For example, local overheating can be detected by ``hot spots'' via high-speed video imaging~\cite{yan2022real,colosimo2018spatially}. Measured quantities like melt pool shapes ~\cite{fox2017measurement} or other process by-products including spatters and plumes are also used in papers~\cite{zhang2018extraction,ye2018situ} to study the thermal history indirectly.

In this paper, we use the melt pool shape as our experimental data for validation. The shape of the melt pool is a direct result of the thermal history. For example, high energy input or high residual heat caused by nearby scanning might lead to a larger melt pool shape. The melt pool length and width reflect the anisotropicity of the thermal history. By analyzing the melt pool shape, it is possible to gain insight into the thermal history of the process and make adjustments to optimize the resulting product. For example, the melt pool shape has been used to predict overheating or underheating thermal defects in paper~\cite{moran2021scan}.

The melt pool shape data we used is generated by the Additive Manufacturing Metrology Testbed (AMMT).} AMMT is a fully custom, open-platform laser powder bed fusion system built by the National Institute of Standards and Technology (NIST)~\cite{lane2019process,fisher2018toward}. This platform uses co-axial melt pool monitoring (MPM) camera, which is capable of taking high-speed high-resolution melt pool frames \cite{lane2020process}. Much research has been conducted with MPM images. For example, Zhang et al. ~\cite{zhang2020neighborhood} built a data-driven melt pool prediction model using AMMT data to keep the melt pool size as constant as possible. Ho et al. ~\cite{yeung2018implementation, yeung2019part} present a new laser power control algorithm that scales the laser power to reduce the variability of melt pool intensity measured throughout the 3D build based on AMMT melt pool data. These papers demonstrated that the melt pool frames can provide highly valuable information even without calibrated temperature distribution.

\subsection{Numerical simulation of LPBF thermal history}

Alternatively, computer simulation can be used to predict LPBF thermal history. Generally speaking, simulation approaches can be categorized into micro-level, path-level, and part-level simulation. The micro-level simulation requires high-fidelity modeling of multiphysics including powder dynamics, fluid dynamics of molten metal, and the interaction between molten metals and metal powders, to list a few~\cite{michopoulos2018multiphysics}. The micro-level simulation, therefore, is computationally expensive and often limited to a short printing path. The finite element method (FEM) is a common approach ~\cite{yan2015multiscale, dunbar2016experimental, cao2021novel} to simulate the path level LPBF process. Compared to micro-level simulation, FEM-based approaches simplify the complex multiphysics problem inside the melt pool. However, a naive implementation of FEM is still computationally expensive because it not only requires small time steps to resolve the high cooling rate at more than 100 °C/s ~\cite{acharya2017prediction}, \ch{but also maintains a high spatial resolution to capture large size difference between melt pools (~100 um) and printed part (~10-100 mm). }

The need for high discretization resolution, small time steps, and the associated high computational costs prohibits the broader adoption of FEM-based approaches in practical engineering applications. For example, only a single-scan path and simple scan strategies are simulated on the path level due to the high computational cost in ~\cite{bruna2018selective} and~\cite{bayat2019keyhole}, respectively. In comparison, a 2 mm $\times$ 2 mm 3-layer structure may require up to 10 hours to simulate~\cite{cao2021novel}. As a result, many consider path-level simulations too computationally expensive to be practical for engineering applications ~\cite{bartlett2019overview, gouge2019experimental}. Multiscale and agglomeration techniques, such as inherent strain~\cite{liang2018modified} and flash heating~\cite{bayat2020part,zhang2019resolution}, have been proposed to bypass the computational complexity in the part-level thermal history simulation. \ch{In these methods, the thermal history on the path-scale level is assumed to be uniform so that it can be “lumped” or “agglomerated” on the layers. The loss of path-level information in thermal history, in turn, leads to inaccurate downstream simulations of, warping, residual stress, material properties, and the overall performance of the part ~\cite{liu2016homogenization,gouge2019experimental}, to list a few.}  Graph theory~\cite{yavari2019thermal} is introduced to make a more efficient simulation of the LPBF process. Adaptive mesh~\cite{ganeriwala2021towards} is also introduced to alleviate the computational cost. The effect of scanning strategies in the LPBF process is investigated in papers~\cite{liu2022understanding, nadammal2021critical, chen2021prediction}, where post-process measurements such as stresses are used to validate their approaches. \ch{In summary, the state-of-the-art thermal history simulation approaches are either too computationally expensive or not accurate enough due to unaccounted assumptions.}

\subsection{Overview of CAPL}
\label{sec:CAPL}

A Contact-aware path-level (CAPL) discretization approach was recently developed for simulating the path-level thermal history of a moving heat source ~\cite{zhang2018linear,zhang2019towards,zhang2022scalable}. The CAPL approach was proposed to improve the stat-of-the-art in two aspects: (1) \ch{A discretization base on the manufacturing tool path}, and (2) a localized thermal history computation based on the contact graph. CAPL has been shown to outperform adaptive remeshing methods ~\cite{olleak2020part} in FEM simulation since the remeshing step is very expensive ~\cite{zhang2022scalable}. 

\begin{figure}
    \centering
    \includegraphics[width=0.45\textwidth]{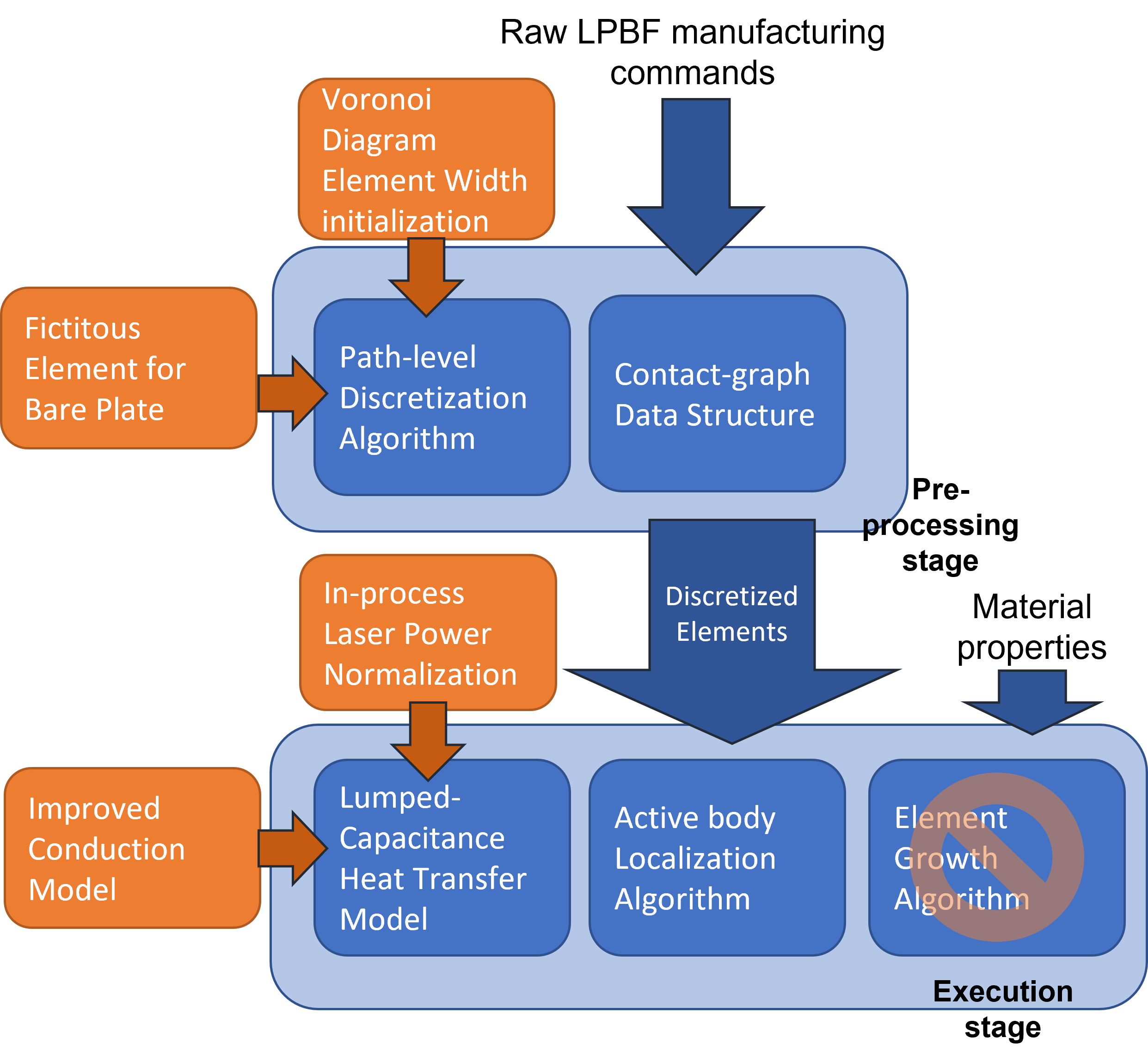}
    \caption{System diagram of contact-aware path-level (CAPL) for laser powder bed fusion (LPBF) process. The components of the original CAPL are in dark blue and our modification and improvements for validation are shown in orange.}
    \label{fig:CAPL_diagram}
\end{figure}

The CAPL approach was proposed for thermal simulation of the additive manufacturing process driven by a moving heat source, including fused deposition modeling (FDM)~\cite{zhang2018linear} and powder bed fusion (PBF)~\cite{zhang2019towards,zhang2022scalable}. CAPL consists of a pre-processing stage and an execution stage with multiple components (see Figure \ref{fig:CAPL_diagram}). In the pre-processing stage, elements are generated by path-level discretization. Then the contact-graph data structure is initialized. In the execution stage, simulation based on the lumped-capacitance heat transfer model assisted by the contact graph and elements. The active body algorithm is used to achieve linear time complexity. The element width and contact graph are updated by the element growth algorithm. For the sake of completeness, we include the following description of components that are relevant to the current work. The interested reader may refer \cite{zhang2022scalable} for details.

In the path-level discretization algorithm, the scan path and laser power information are extracted from the input file and discretized in both space and time into some sub-paths. CAPL considers an element to be the material that has been newly melted/sintered by a laser scan along the corresponding sub-path, so each element must correspond to a sub-path with positive laser power. Each element is approximated as a \ch{path-aligned} box defined by its length $L$ (equal to the length of the sub-path), width $W$, and height $H$ (equal to the layer height). The cross-sectional shape is a rectangle $W \times H$ which is perpendicular to the scanning direction. The top surface of the element is approximated by a rectangle $L \times H$. The initial width of an element $W_0$ is predefined. This width will be dynamically grown in the execution stage assisted by the element growth mechanism. The original CAPL approach requires no overlap between different top surfaces of different elements. The connectivity (adjacency information between elements), which is represented by a spatial structure called a contact graph, is initialized with the initialized element positions and sizes. The previous CAPL implementation only supported orthogonal scan paths as it assumed that the elements are aligned. 

The element growth mechanism is introduced in the simulation to mimic the powder melting process. Each element is initialized with a small initial width ($W_0$) which is determined by empirical data or high-fidelity simulations. The initial width is always smaller than the final width because thermal expansion will only enlarge the element width. Element width will grow according to its thermal conditions via an iterative correction procedure in the execution stage. The width growth is triggered when an element’s temperature is higher than some predefined threshold which mimics the melting of the element and its thermal width expansion. 

The lumped-capacitance heat transfer model is a lumped parameter model which is derived from the energy conservation of laser energy input $H$, radiation $Q$, convection $Q_{conv}$, and conduction $Q_{cond}$. The lumped parameter models have been commonly used in many engineering systems~\cite{wang2019topological} to represent the spatially and temporally distributed physical phenomena. The laser heating term $H$ only applies to the elements in the top layer. $H$ is integrated on the top surface of the element by assuming that the laser beam has an ideal Gaussian intensity profile and a Beer-Lambert type model is used where laser intensity decreases exponentially with respect to the penetration depth. The convective term $Q_{conv}$ is used to represent the energy dissipation from a free surface to the ambient environment through heat convection. The conductive term $Q_{cond}$ includes (1) conduction to neighboring elements along the scanning direction, (2) conduction to build the platform, and (3) conduction to contacting elements in adjacent paths \ch{, and (4) conduction to the neighboring layers and powders}. The heat conduction is driven by the temperature gradient \ch{which} is approximated by the finite difference method. The temperature of the element is assumed to be uniform within each element since it is assumed that the Biot number of the element is small, in other words, the element is assumed to be small enough so that it is determined to be ``thermally simple''. Readers can refer to paper \cite{zhang2022scalable} for more details.

\section{Modifications and improvements on Contact-Aware Path-Level (CAPL) thermal simulation}
\label{sec:CAPLlimit}


\q{CAPL originally focused on the FDM process and made corresponding assumptions.} Modification is needed to use CAPL to simulate the laser scanning on the solid surface. 
In this section, we first identify the \ch{modifications} that are needed for CAPL to be consistent with implementation details in the NIST AMMT dataset. We then discuss the necessary modifications and improvements to address these discrepancies. (see Figure \ref{fig:CAPL_diagram}).

To attempt to improve the simulation results and better utilize and align with available data, \ch{We identified three needed modifications to be consistent with} implementation details in the NIST AMMT dataset. Firstly, the original CAPL requires all the elements must be associated with a segment of scanning paths. The elements represent the powder materials that will be scanned and solidified. The widths of these elements will be dynamically grown in the execution stage to mimic the powder melting process. However, the AMMT data used is obtained on a solid surface. Because it is a continuum all the volume needs to be discretized and accounted for in the simulation no matter if scanning paths go through it or not. Also, since there is no powder melting process on the solid surface, the element growth is not applicable on the base layer. 

Secondly, the original CAPL approach requires the element to have no overlap with other elements. The element \ch{contacts needs to be orthogonal since all elements have a similar rectangular shape.}
\q{But in the PBF process, it is common to have overlapping paths since it is possible to remelt the solidified powder. In addition, the scan strategy used for the NIST experiment has nonparallel paths (see contour and infill in Figure \ref{fig:fictitious paths}).} Unexpected overlap or void will occur when addressing the general non-parallel paths, as shown in Figure \ref{fig:voronoi_elements}.

Lastly, the original CAPL uses the forward Euler method (a first-order explicit method) for thermal simulation and assumes a small Biot number ($\text{Bi} = hL/k$, where $h,L,k$ is the convection coefficient, characteristic length, and conductivity of the element) for the element. In the FDM process, elements usually can satisfy this assumption due to the scanning speed being relatively slow compared to the PBF process. Here, however, a small element length ($\sim$10 $\mu$m) is required to ensure enough resolution to capture the melt pool length, but the element width is too large since the hatch space is too large (100 um here). This means the Biot number in the width direction will be an order larger than that in the length direction and should be reflected as such in a PBF model. In the lumped model, thermal conduction between two elements $i$ and $j$ is computed as $Q_{cond} = kA\Delta T_{ij}/d_{ij}$ where $k$ is the conductivity, $A_{contact}$ is the conduction area, and $d_{ij}, and \Delta T_{ij}$ are the distance and temperature difference between the two elements. Considering the elements here (see Figure \ref{fig:voronoi_elements}), a naive implementation of the lumped model will lead to insufficient heat conduction in the direction where the characteristic length is too long. 


\subsection{Representing solid surface with fictitious elements}

\begin{figure}
    \centering
    \includegraphics[width=0.4\textwidth]{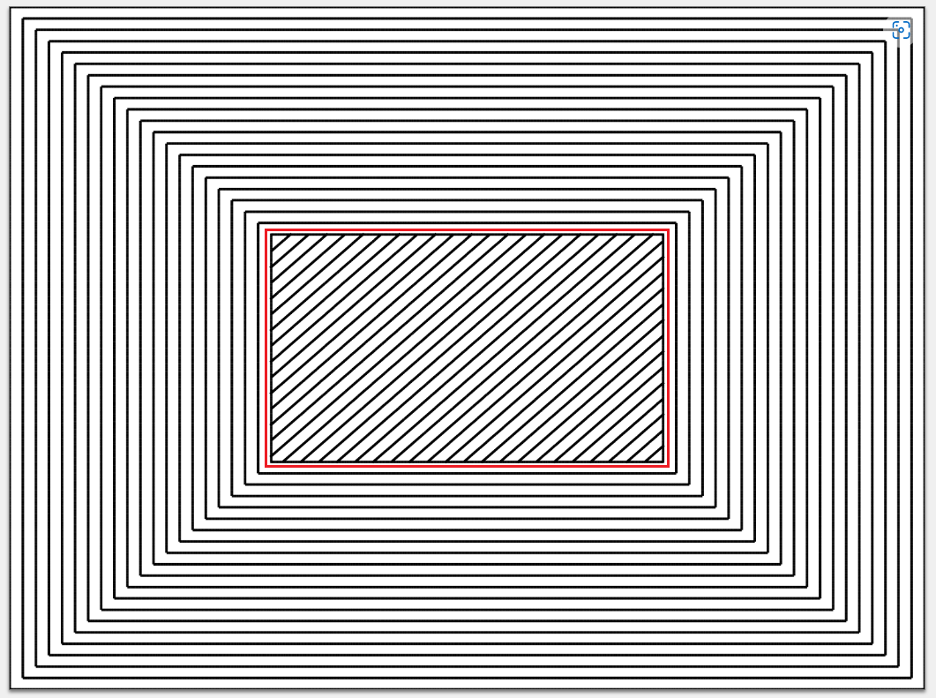}
    \caption{The actual laser path is inside the red box. Fictitious elements are added as the paths outside the red box to represent the larger surface. The solid continuum is modeled as multiple such layers.}
    \label{fig:fictitious paths}
\end{figure}

For the first \ch{improvement}, we modify the path-level discretization algorithm in the pre-processing stage to account for solid continuums. We add additional fictitious paths so that the entire domain is filled with real paths and fictitious paths. The contacts between all elements are initialized according to the contact graph. We initialize all elements with solid material properties since a solid surface is used in the present paper. Since no powder melting is involved in the present paper, we suppress the element growth mechanism. For the present paper, the fictitious paths are the contours around the real scanning paths, as shown in Figure \ref{fig:fictitious paths}. 

\subsection{Elements initialization by Voronoi diagram}

\begin{figure}
    \centering
    \includegraphics[width=0.4\textwidth]{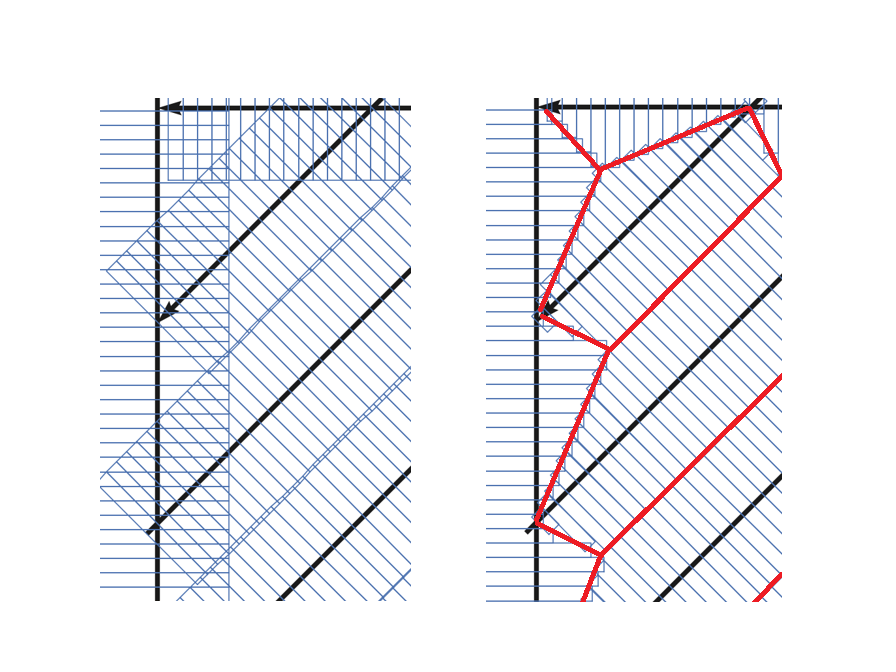}
    \caption{Example of elements on the left top corner of the layer.  Unexpected element overlaps with non-parallel overlapped path exists before modification (left). Elements generated by Voronoi diagram after modification (right): elements (in blue) approximately form a Voronoi diagram (in red) whose sites are the scanning paths (in black)}
    \label{fig:voronoi_elements}
\end{figure}

To address the second improvement, we modify the path-level discretization algorithm with a new elements width initialization algorithm by the Voronoi diagram. The element width will be no longer updated in the execution stage since we suppress the element growth mechanism as discussed above. In the new element width initialization algorithm, we initialized all element widths with a small number ($W_0 = 10 \mu m$). After initialization, we increase the element widths simultaneously. For each element, we stop the increase of width once the element's overlap with other elements exceeds a given threshold $s$ ($3\times 10^{-11} m^2$ used here). The element width initialization ends when all elements stop growing wider. Such element width initialization procedure approximately generates a Voronoi diagram, whose cells have the path segments as their site, see Figure \ref{fig:voronoi_elements}.

\ch{We use an }in-process laser power normalization for the second \ch{needed modification} in addition to the element width initialization by the Voronoi diagram. Note there are some overlaps and voids between the elements initialized by the Voronoi diagram. When a laser scans through these elements, the voids and overlaps will cause the total laser power input to artificially fluctuate. The in-process laser power normalization is a modification of the lumped-capacitance heat transfer model \ch{to correct for this artificial fluctuation}. In the lumped model, every element takes the laser energy input\ch{. The} total laser energy input on all elements should be equal to $\alpha P$ (laser power $P$ multiplies absorptivity $\alpha$) because of energy conservation:
\begin{align}\label{eq:H}
\sum_i^N H_i(t) = \alpha P
\end{align}
where $H_i(t)$ is the laser energy input term of the element $i$. \ch{Theoretically $N$ should be the number of all elements which are exposed to laser energy input. Here, we use the number of elements inside the active body as $N$. This approximation is due to the assumption of thermal localization by utilizing an active body.} The total energy could be higher or lower than the input laser term because of the overlap and void. To mitigate this problem, we normalize the input laser power $P$ during the simulation process by a constant $\alpha P/\sum_{i}^{N} H_i$. In other words, the laser power at $t$ should be:
\begin{align}
P(t) =  \frac{\alpha P}{\sum_{i}^{N} H_i} P
\end{align}
in which $P$ is the nominal laser power, and $P(t)$ is the normalized laser power to replace $P$ in equation \ref{eq:H}.  

\subsection{Improved conduction model}
\begin{figure}
    \centering
    \includegraphics[width = 0.2\textwidth]{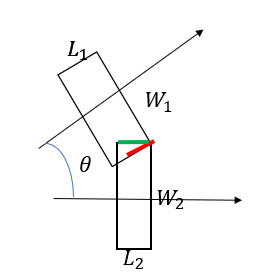}
    \caption{Top view of the contact area (in red). The contact area is the smaller projected cross-section (in green) projected along the angle $\theta$ between two elements. $L$ and $W$ are the element length and width. Arrows indicate the scanning directions.}
    \label{fig:cs}
\end{figure}

The third improvement addresses the lumped thermal model for the elements. \ch{Here we correct a previously underestimated temperature gradient, which is originally approximated by the finite difference method. For example, the original conduction energy is given as $kA_{contact}\frac{\Delta T_{ij}}{d_{ij}}$, where the conduction characteristic distance is $d_{ij}$ which is the distance between two elements $i$ and $j$. In a real melt pool, the temperature spatial gradient is very high around the melt pool (temperature spatially decays from the melting point to a much lower temperature in a relatively short distance). When the element width is too large, the same temperature decay will be artificially assumed to happen at a much larger distance (the element width), thus the side-by-side conduction will be underestimated. To address this issue,} we cap the conduction characteristic distance by thresholding the conduction characteristic distance. \ch{Here we assume the threshold $d_0$ to be a constant which equals to} the maximal element length (here is 10 um). Since here element length is much smaller than the element width, the threshold $d_0$ ensures the thermal conduction $Q_{cond}$ in both the scanning direction and the transverse direction has the same order of magnitude. The conduction term between two elements $i$ and $j$ is to be modified as 
\begin{align}
Q_{cond} = \frac{kA_{contact}\Delta T_{ij}}{\max(d_0,d_{ij})}
\end{align}
\ch{Due to the existence of overlapped elements, we redefine the contact area} as the projection area (the projected angle is $\theta$) of the intersected cross-section (see Figure \ref{fig:cs}):
\begin{align}
    A_{contact} = \min(A_{cs1},A_{cs2})/\sin{\theta}
\end{align}
where $A_{cs} = W\times H$

\section{Validation of PBF-CAPL through Melt Pool shapes}
\label{sec:validation}

\subsection{Validation Dataset Overview}

\begin{figure}
    \centering
    \includegraphics[width=0.4\textwidth]{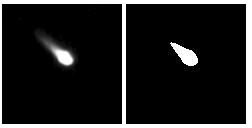}
    \caption{An example of melt pool frame (left) acquired on the Additive Manufacturing Metrology Testbed by NIST and the binarized result (threshold is 80 out of 255).}
    \label{fig:melt_pool_example}
\end{figure}

The experiment dataset used for validation of the modified CAPL consists of melt pool frames acquired on the Additive Manufacturing Metrology Testbed built by NIST. The resolution of each frame is approximately 7.13 um/pixel. An example of the melt pool frame is shown in Figure \ref{fig:melt_pool_example}. A threshold of the digital value corresponding to the melting temperature is needed to extract the melt pool shape. Based on other experiments conducted on AMMT~\cite{yeung2020residual}, we use 80 out of 255 as the threshold (see Figure \ref{fig:melt_pool_example}). \ch{The melt pool shape is approximated by the Python library scikit-image's skimage.measure.regionprops function (equivalent to MATLAB’s regionprops function)}, which approximates the melt pool shape by an ellipse that has the same normalized second central moments as the region. The melt pool length and width are computed as the length of the major and minor axis.

\begin{figure}
    \centering
    \includegraphics[width=0.4\textwidth]{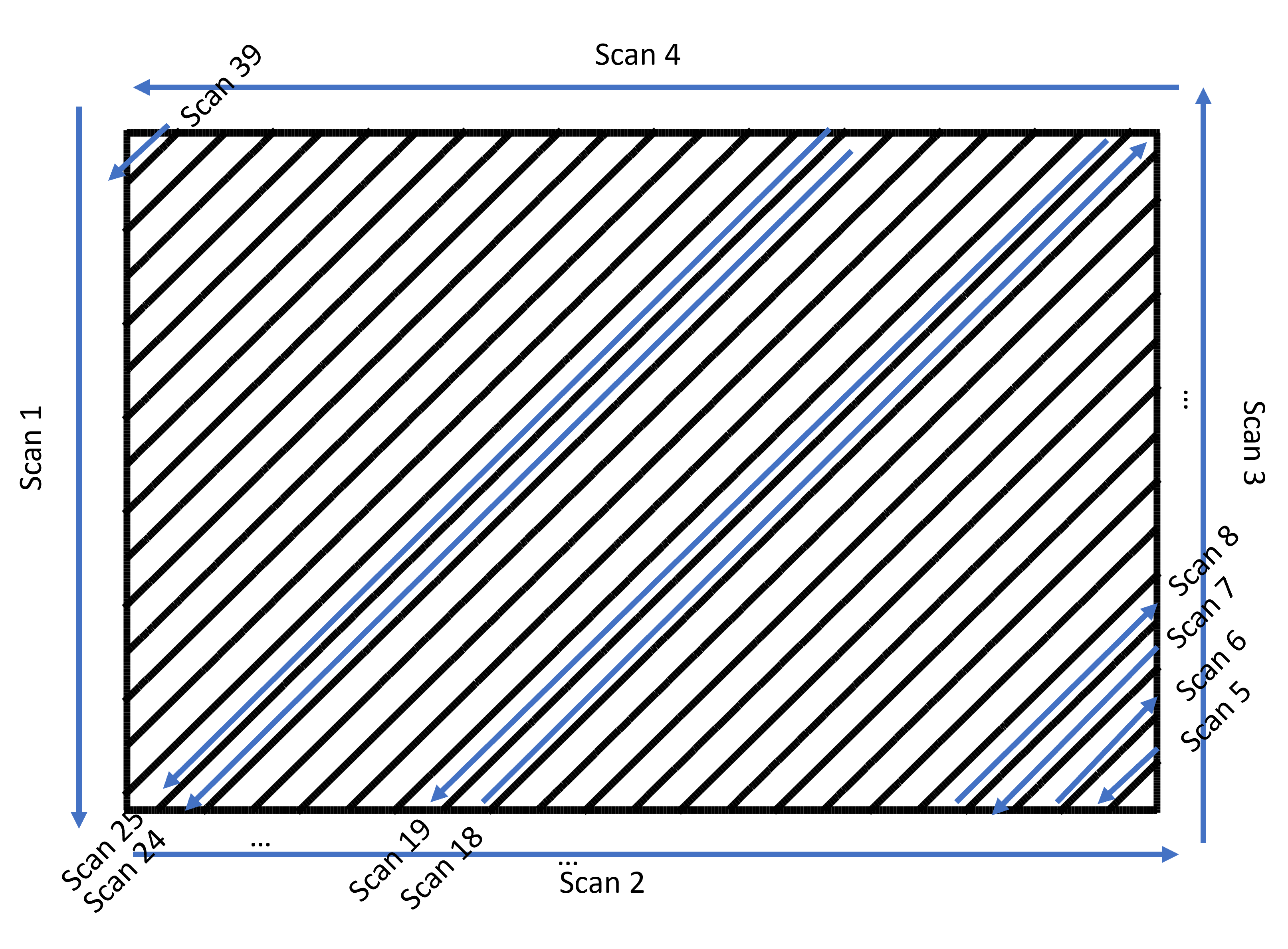}
    \caption{Numbering of scan vectors.}
    \label{fig:scanid}
\end{figure}

\begin{figure}
    \centering
    \includegraphics[width=0.45\textwidth]{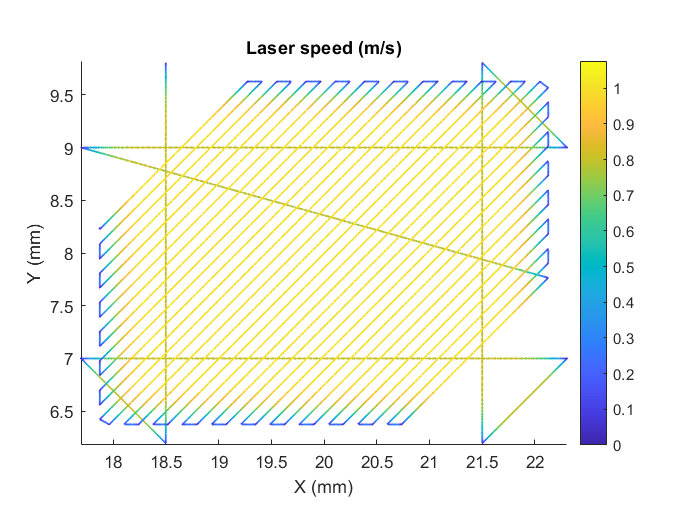}
    \caption{Laser speed map for Case 01. All 10 cases have the same identical speed map only different by their relative locations. Only Case 01 has constant laser power.}
    \label{fig:laser_speed_map}
\end{figure}

\begin{figure}
    \centering
    \includegraphics[width=0.47\textwidth]{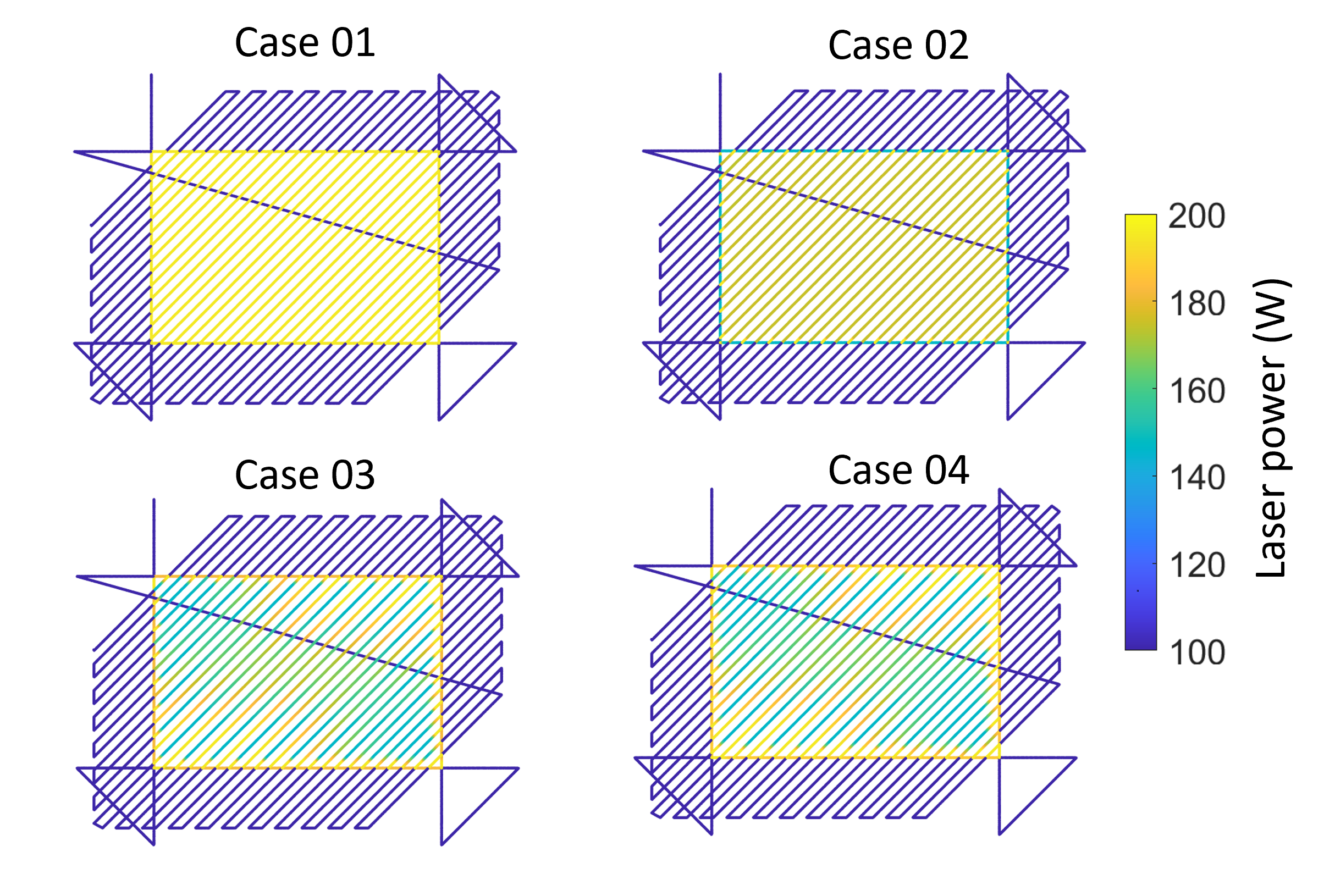}
    \caption{Laser power map for cases 01 - 04. The dark blue here indicates no power. From Case 02, all cases have variant laser power and they differentiate from each other by how the laser power distributes.}
    \label{fig:laser_power_map}
\end{figure}

\begin{figure}
    \centering
    \includegraphics[width = 0.4\textwidth]{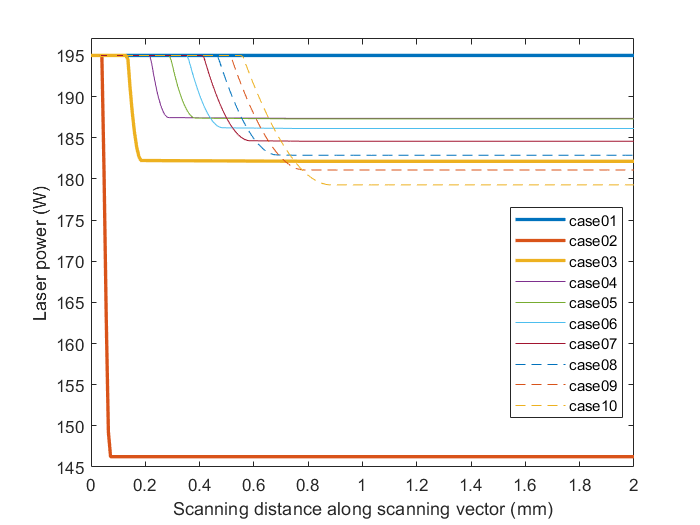}
    \includegraphics[width = 0.4\textwidth]{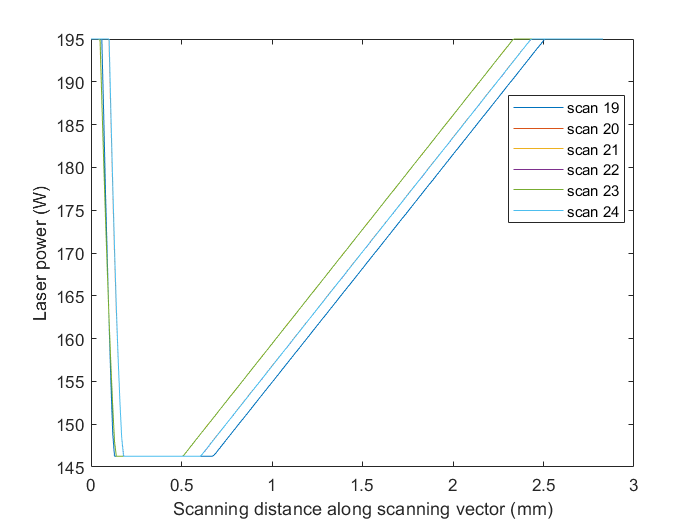}
    \caption{Laser power plotted as a function of distance along the Scan 01 for all cases (top). The distance range from 0 to 2 mm because the length of Scan 01 is 2 mm for all cases. The laser power of scan 19 - 24 for case 03 is shown at the bottom. Scan 19 - 24 have adjacent parallel scan vectors which have the same length.}
    \label{fig:laser_power_cases01_10_scan01}
\end{figure}

We use the first 10 cases from the dataset with varying laser power and every case consists of 1498 melt pool images at various locations during the laser scan. The scanning paths are shown in Figure \ref{fig:scanid} and all 10 cases use the same scanning path. There are in total 39 scan vectors which are labeled in Figure \ref{fig:scanid}. \ch{The laser scanning speed for all 10 cases is the same and is shown in Figure \ref{fig:laser_speed_map}. Note the laser is turned off while overshooting outside the scanning region to ensure constant scanning speed inside the scanning region.} For ease of discussion, we classify scan vectors into different groups, including 4 contour vectors and 35 diagonal raster scan vectors inside the contours (see Figure \ref{fig:scanid}). The 10 cases differentiate from each other by the laser power being used (see Figure \ref{fig:laser_power_map}). Case 01 has the highest laser power which is constant (195 W), and all other cases have variable and lower laser power. For example, the laser power of Scan 01 for all cases and the laser power of Scan 19 - Scan 24 for Case 03 are shown in Figure \ref{fig:laser_power_cases01_10_scan01}, in which the laser power is plotted as a function of the distance along the scanning direction. The laser power was lowered to decrease the variance in melt pool size based on the residual heat factor in an earlier NIST study. Interested readers may refer \cite{yeung2020residual} for details. For every case, the 1498 melt pool images come from sampling on a single-layer scanning on an Inconel 625 solid surface. The images come from an even sampling at the frequency 20kHz when the laser is turned on. 

\begin{figure*}
    \centering
    \includegraphics[width=0.9\textwidth]{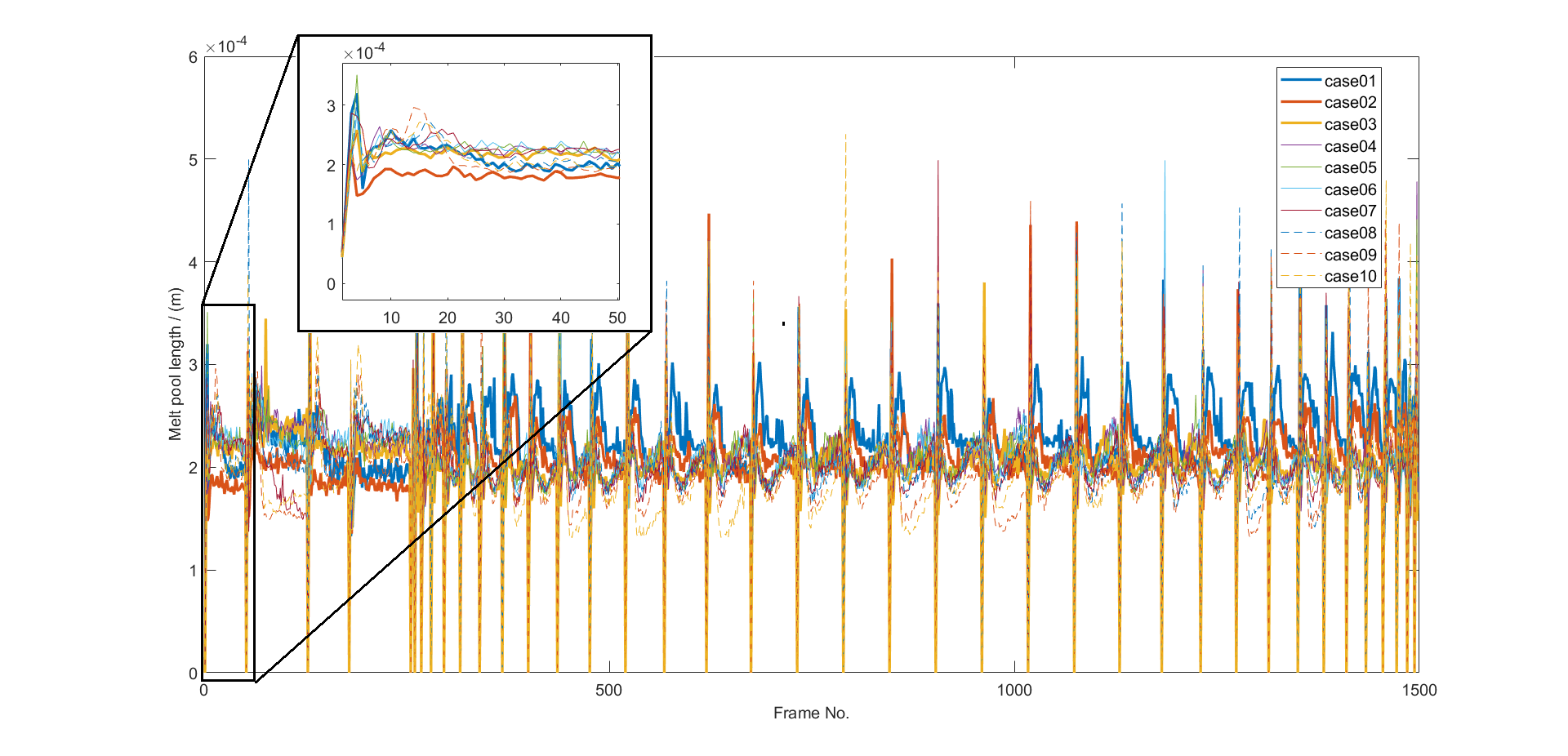}
    \caption{Experimental melt pool length for all 10 cases $\times$ 1498 frames. Zoom in view of Scan 01 is shown. Zero melt pool length is due to the MPM system skipping the first frames of the scan vectors.}
    \label{fig:all_cases_1498_length_NIST}
\end{figure*}

All melt pool lengths from the experimental melt pool image data are plotted as line graphs shown in Figure \ref{fig:all_cases_1498_length_NIST}. The vertical axis represents the melt pool length, and the horizontal axis represents the frame number. Since the scanning path is geometrically the same for all cases and the frames are sampled at the same time sequence, the same frame number results in the same relative location in every case.

\begin{figure}
    \centering
    \includegraphics[width = 0.45\textwidth]{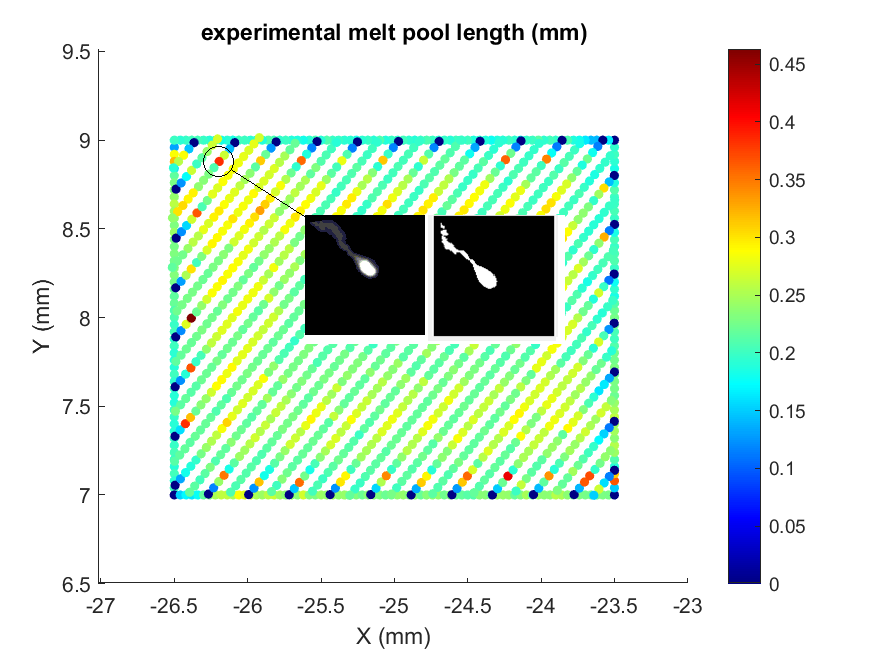}
    \caption{Melt pool image with plume and the binarized results. Plume leads to incorrect melt pool length extracted value. These incorrect values are shown in the length map as red dots.}
    \label{fig:plume_AMMT_frames}
\end{figure}

In all cases, we observe that the melt pool lengths periodically drop to zero. This is due to a synchronization issue of the melt pool monitoring system such that the first image at the beginning of each scan is captured before the laser starts. In addition, excessively long melt pools are observed near the beginning of some scan vectors. Upon closer examination, this is due to the gas plume being mischaracterized as the melt pool after thresholding (see Figure \ref{fig:plume_AMMT_frames}). In both these cases, the extracted melt pool lengths do not correspond to the real melt pool length, therefore, are considered outliers in the CAPL validation. 

\subsection{Calibration of process parameters and material properties}

Melt pool length from CAPL simulation is directly computed by measuring the maximum distance between two melted elements. We use all material properties but absorptivity from \cite{arisoy2019modeling} for simulations in the present paper. Material properties including specific heat and conductivity are modeled as functions of temperature (see table \ref{table:in625}). Phase changes due to melting or solidification are handled by the equivalent specific heat formulation \cite{arisoy2019modeling,zhang2020scalable}.  

The absorptivity is calibrated with Case 01 of the dataset and used for the remaining cases. We calibrated absorptivity because it has a significant influence on thermal history and it can significantly vary when the surface conditions vary. Understanding absorptivity and surface conditions itself can be an important topic where many studies have been conducted \cite{trapp2017situ,matthews2018direct}. We note that the absorptivity of laser power may depend on a variety of factors and is not a constant in general, but it is a common practice that constant absorptivity is used for simplification. In the present paper, we choose 0.43 as it minimizes the error of melt pool length in Case 01. This value is consistent with the ranges provided by the literature. For example, the absorptivity of In625 solid surface ranges from 0.4 to 0.9 \cite{lane2020transient}. This value is applied to other cases in the subsequent tests. 

\begin{figure}
    \centering
    \includegraphics[width=0.47\textwidth]{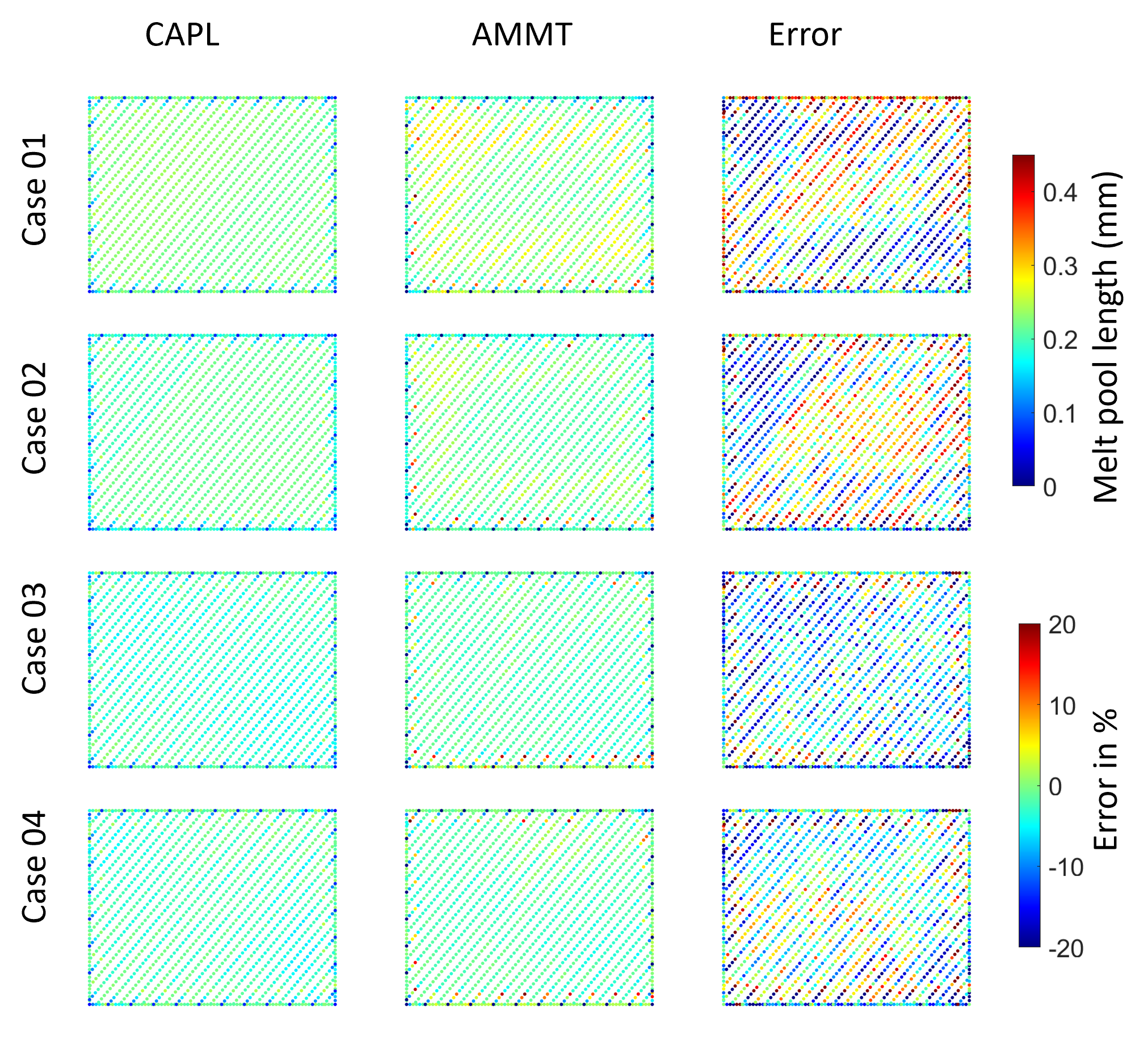}
    \caption{AMMT experimental results of melt pool length map for cases 01, 02, 03, and 04 (middle) and the corresponding CAPL simulation results (left) and the relative errors in percentage (right). AMMT and CAPL results share the same color bar.}
    \label{fig:NIST_lengths}
\end{figure}

Experimental data and simulation results of melt pool lengths, as well as the error percentages, are plotted in Figure \ref{fig:NIST_lengths}. We observe a consistent match between the experimental data and simulation results with about 10 percent of relative error in all cases as shown in table \ref{table:err}. The comparison between simulation results and experimental data will be discussed in detail in the rest of this section.

\begin{table}
\caption{Process parameters and material properties of IN625\cite{arisoy2019modeling}}\label{table:in625} 
\centering

\footnotesize
\begin{tabular}{ll}
 \hline
 Solidus temperature $T_S$ (K) &   1563  \\
 Liquidus temperature $T_L$ (K)&   1623 \\
 Environmental temperature (K)&   293 \\
 Latent heat of fusion (kJ/kg) & 290  \\
 Specific heat ($T \leq T_S$) (J/kg K) & 339 + 0.24T  \\
 Specific heat ($T \leq T_L$) (J/kg K) & 735  \\
 Thermal conductivity, ($T \leq T_S$) (W/(m K))& 5.3 + 0.015T \\ 
 Thermal conductivity, ($T \leq T_S$) (W/(m K)) & 30.05 \\
 Density kg/m$^3$ & 8440  \\
 Convection coefficient (W/(m$^2$ K)) & 10 \\
 Substrate temperature (K) & 293  \\
 Laser spot diameter ($\mu$ m) & 85
\end{tabular}
\end{table} 

\subsection{Melt pool length validation}


\subsubsection{Melt pool length evolution within scan vector}
Firstly we discuss the melt pool length evolution on the same scan vector.  We categorized the results into two groups: contour scan vector (we will use Scan 01 as an example) and 45-degree scan vector (we will use Scan 19 - 24 as examples). 

\begin{figure}
    \centering
    \includegraphics[width = 0.4\textwidth]{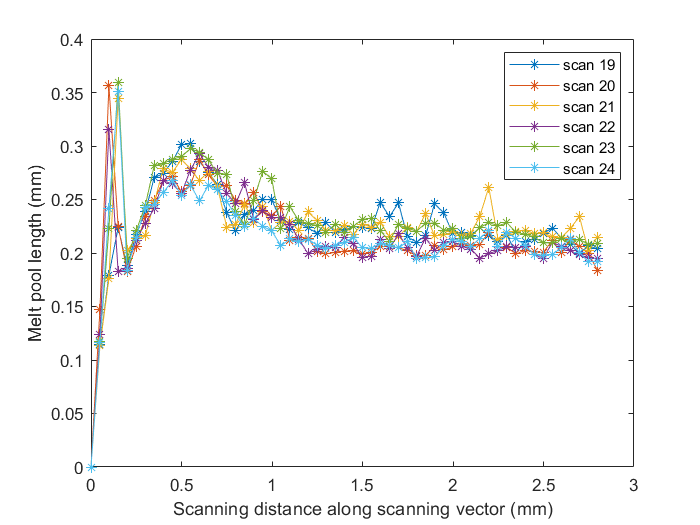}
    \includegraphics[width = 0.4\textwidth]{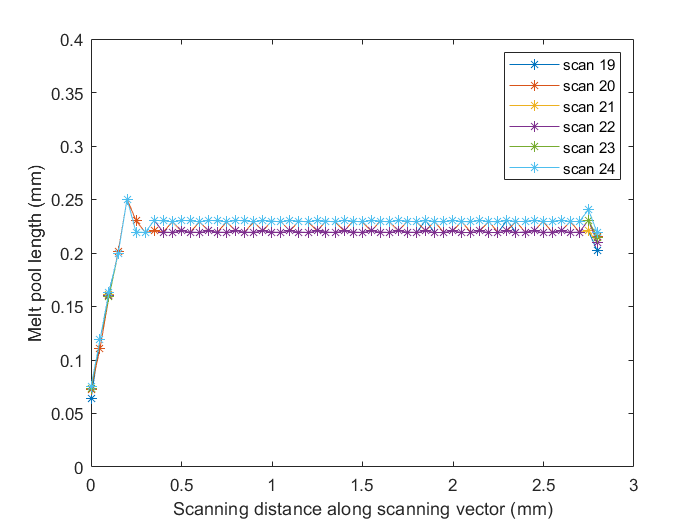}
    \caption{Experimental (top) and simulation (bottom) melt pool length of Scan 19-24 for Case 01. Laser power is constant (195 W). The CAPL approach does not predict the bump.}
    \label{fig:scan19_24_case01}
\end{figure}

We choose Scan 01 since it can be considered a single-scan test without thermal inference from other scan vectors. Because it is the first scan vector and there is no laser scanning on the surface yet before Scan 01 is applied, Scan 01 should be free from the influence of other scan vectors. Therefore the evolving melt pool shape on it should just reflect the influence from Scan 01. We also choose Scan 19 - 24 in Case 01 because (a) they are parallel and identical to each other and all scan vectors have constant laser power, and (b) we observed melt pool length evolves similarly on these scan vectors (see Figure \ref{fig:scan19_24_case01}).

\begin{figure}
    \centering
    \includegraphics[width = 0.4\textwidth]{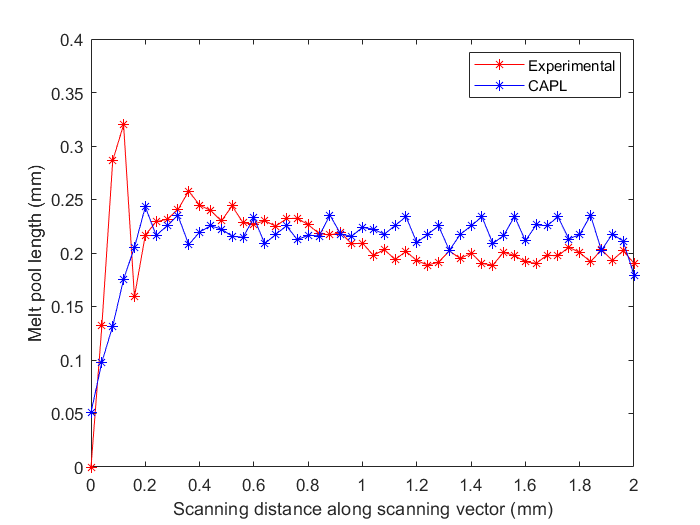}
    \includegraphics[width = 0.4\textwidth]{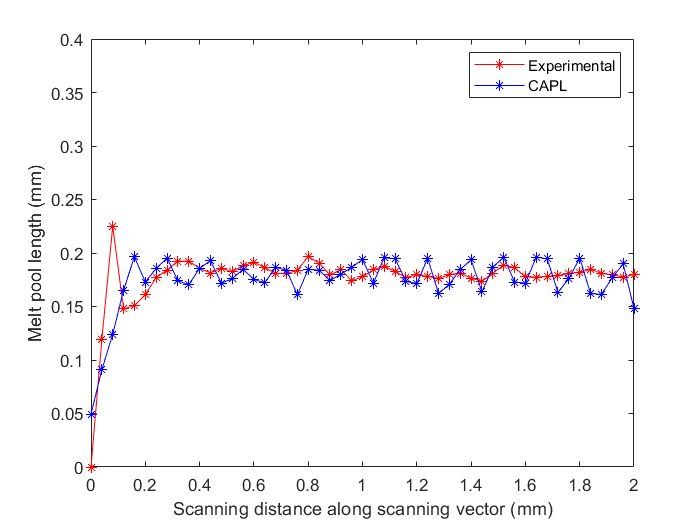}
    \includegraphics[width = 0.4\textwidth]{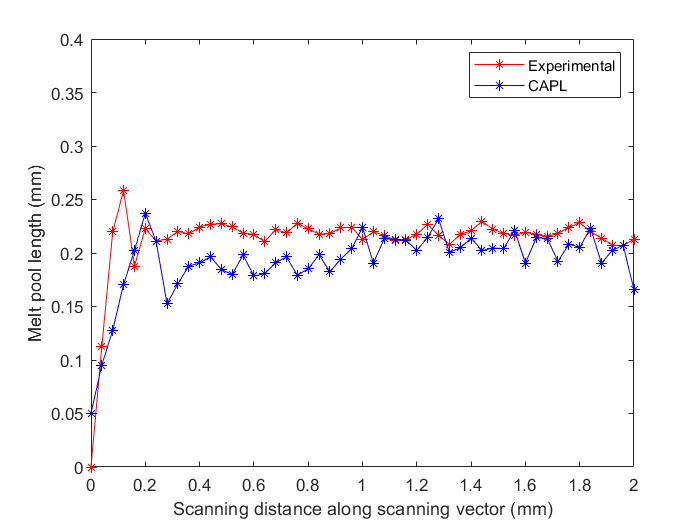}
    \caption{Melt pool length for Scan 01 of Case 01, Case 02, and Case 03. A bump at the beginning of the scan vector in experimental data is not predicted by CAPL in Case 01. Laser power along the scan vector can be found in Figure \ref{fig:laser_power_cases01_10_scan01}.}
    \label{fig:scan01_case0102}
\end{figure}

\begin{figure}
    \centering
    \includegraphics[width = 0.4\textwidth]{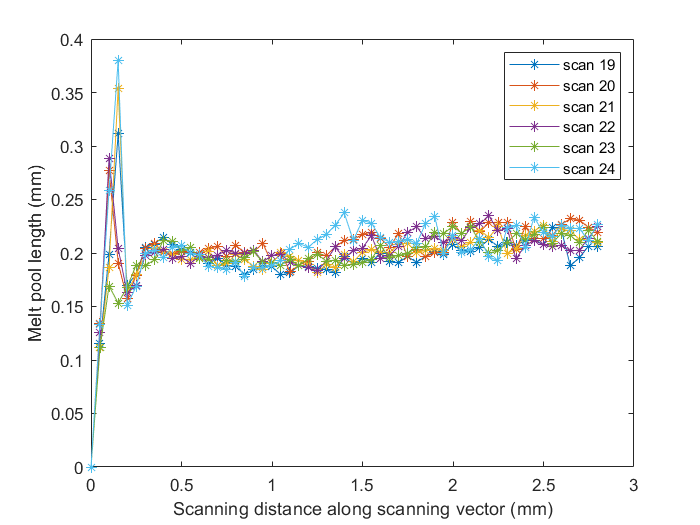}
    \includegraphics[width = 0.4\textwidth]{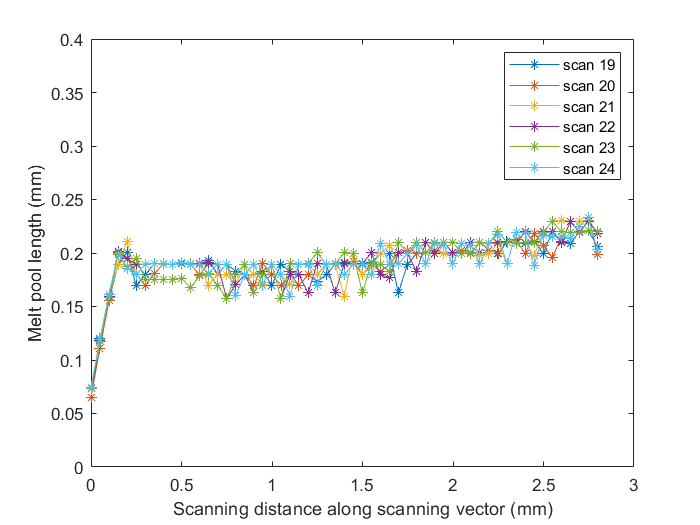}
    \caption{Experimental (top) and simulation results (bottom) of Scan 19 - 24 in Case 03. The bump disappeared in experimental data, and the trend in experimental data is captured by simulation results. The laser power of these scans can be referred to in Figure \ref{fig:laser_power_cases01_10_scan01}.}
    \label{fig:scan19_24_case03_length}
\end{figure}

\begin{figure}
    \centering
    \includegraphics[width = 0.4\textwidth]{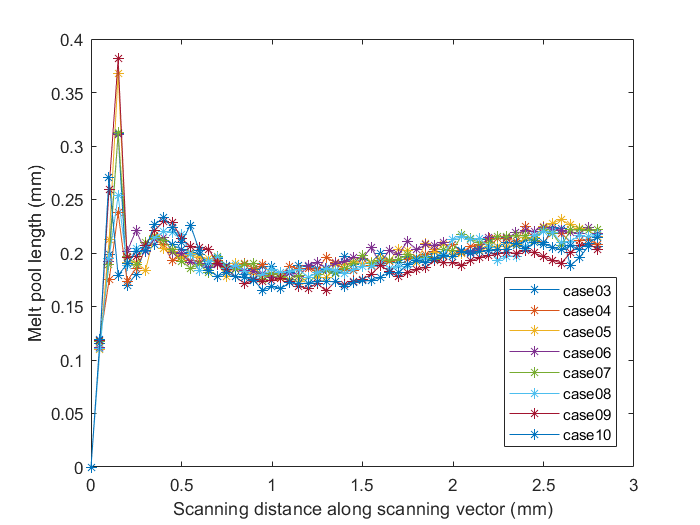}
    \includegraphics[width = 0.4\textwidth]{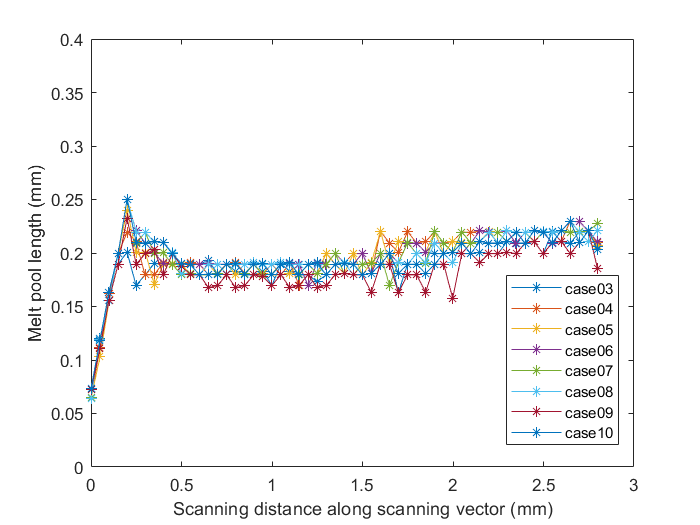}
    \caption{Scan 19 in Case 03 - Case 10 as a function with respect to scan distance. AMMT data (left) vs CAPL simulation results (right).}
    \label{fig:scan19_case03-case10}
\end{figure}

On Scan 01, we plot melt pool length as a function of scanning distance on the scan vector and show the results of cases 01, 02, and 03 in Figure \ref{fig:scan01_case0102}. An increase in melt pool length at the beginning of the laser scan, i.e. ``bump'', is observed in Case 01: melt pool length firstly increases and then decreases as the laser moves on the scan vector.  We also observed the bump on the 45-degree scan vectors (see Figure \ref{fig:scan19_24_case01}) of Case 01. The CAPL simulation does not predict such a bump in these cases. One possible reason for this discrepancy is that we use constant absorptivity in modified CAPL simulation. This is because we observe that the magnitude of the bump is related to the laser power. For example, comparing the melt pool length on Scan 01 in different cases, the bump is much less obvious in cases 02 - 07, where relatively lower power is applied shortly after at the beginnings of the scan vectors. The CAPL simulation matches well with experimental data in these cases (see Figure \ref{fig:scan19_24_case03_length} and Figure \ref{fig:scan19_case03-case10}). We will discuss the increase in melt pool length at the beginning of the laser scan in Section \ref{sec:conclusion}.

\begin{figure}
    \centering
    \includegraphics[width = 0.4\textwidth]{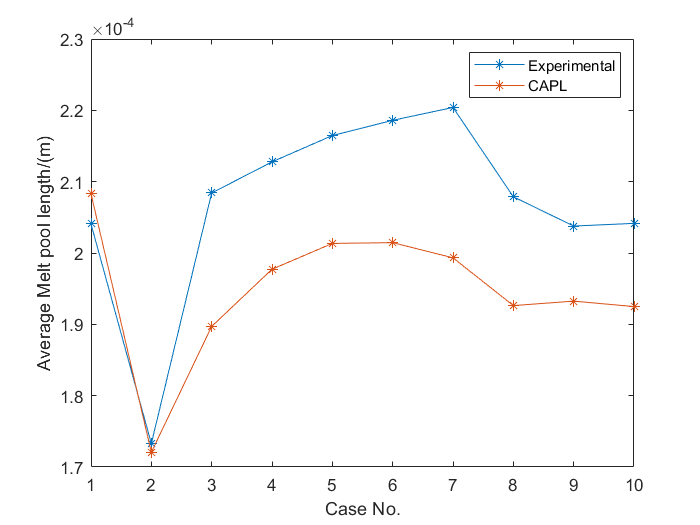}
    \includegraphics[width = 0.4\textwidth]{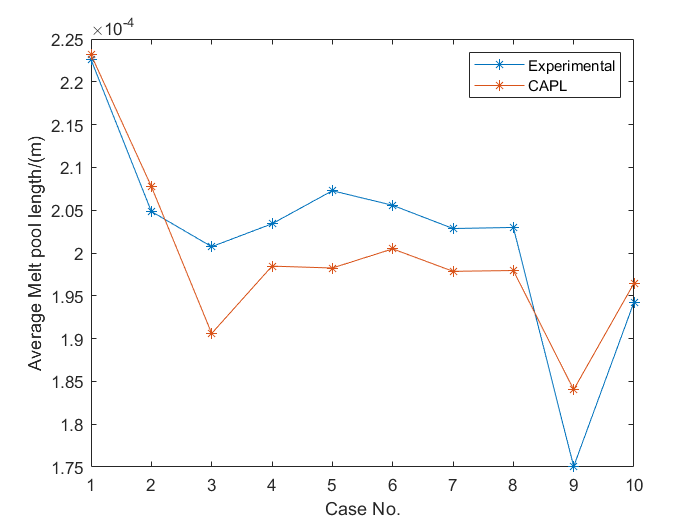}
    \includegraphics[width = 0.4\textwidth]{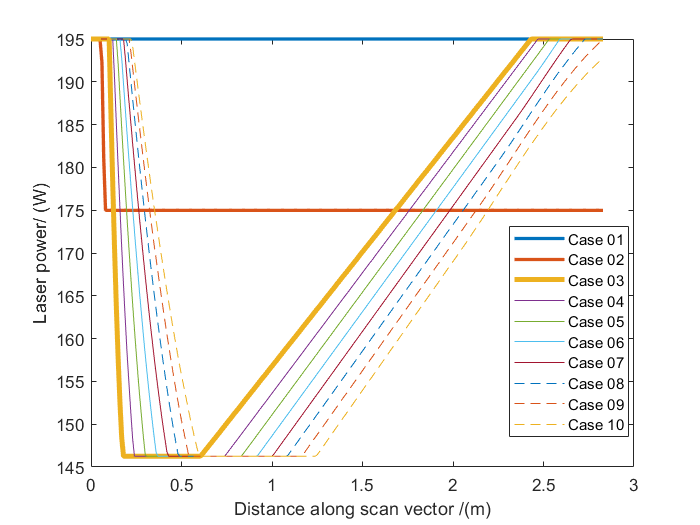}
    \caption{Mean value of melt pool length on Scan 01 (top) and Scan 24 of Case 01 to Case 10 (middle) and laser power on Scan 24 (bottom). CAPL predicts a similar result compared with experimental data.}
    \label{fig:ave_length_scan01_CAPL}
\end{figure}

The effects of varying laser power on the length of the melt pool are also predicted by the CAPL simulation. We give examples in Figure \ref{fig:ave_length_scan01_CAPL}, where the mean values of melt pool length on Scan 01 and Scan 24 are plotted with respect to the case number. We observed that CAPL captures a similar trend as shown in experimental data: on Scan 01, Case 02 has the smallest average melt pool length, and then the average melt pool length increases from Case 03 and then decreases again from Case 07. On Scan 24, both CAPL and experimental data suggest the smallest average melt pool length is in Case 09 and the largest average value is in Case 01.

\subsubsection{Scan-wise melt pool length evolution}

\begin{figure}
    \centering
    \includegraphics[width = 0.4\textwidth]{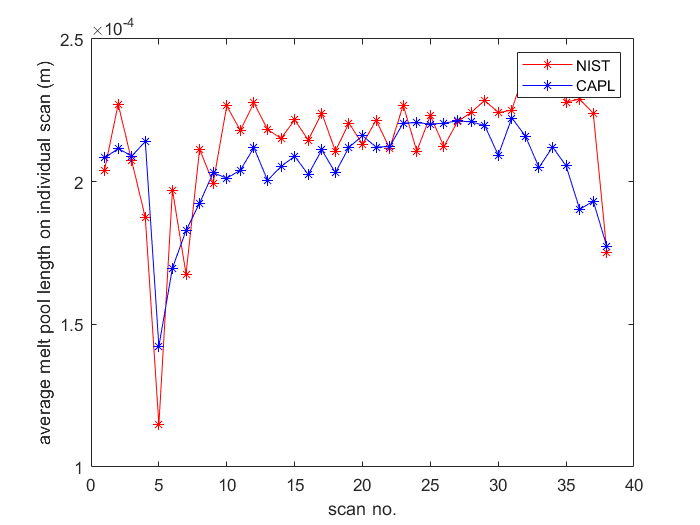}
    \includegraphics[width = 0.4\textwidth]{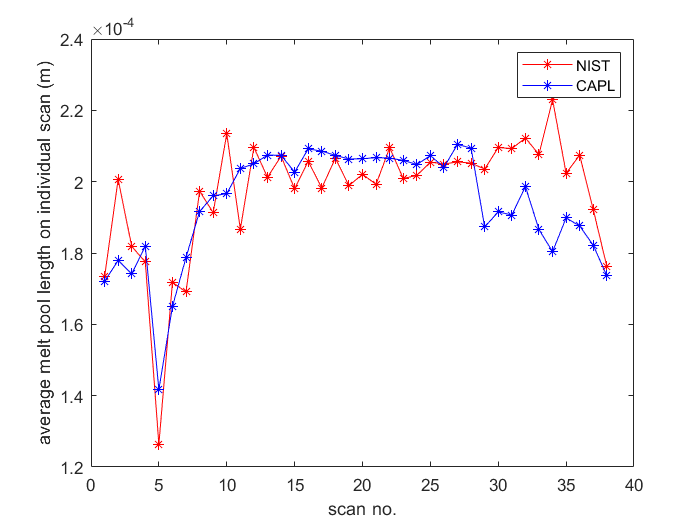}
    \includegraphics[width = 0.4\textwidth]{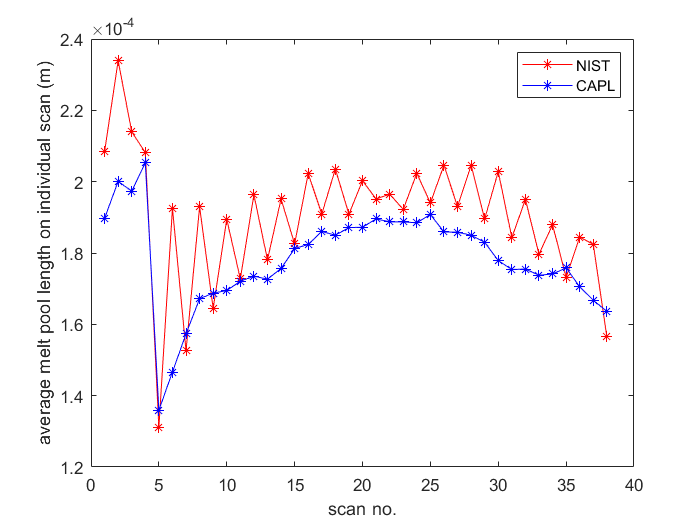}
    \includegraphics[width = 0.4\textwidth]{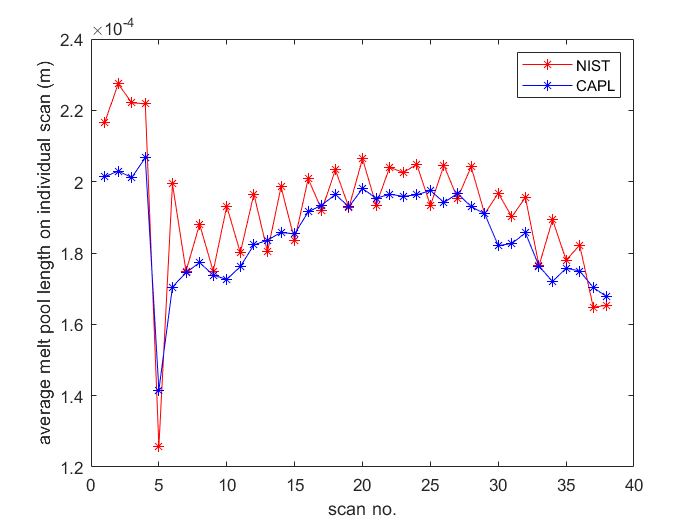}
    \caption{Scan-wise average melt pool length, Case 01, 02, 03, 05. The oscillation of experimental data is due to the perturbation of plume frames.}
    \label{fig:scanwise_ave_length_CAPL}
\end{figure}

\begin{figure}
    \centering
    \includegraphics[width = 0.11\textwidth]{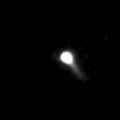}
    \includegraphics[width = 0.11\textwidth]{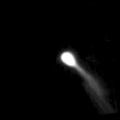}
    \includegraphics[width = 0.11\textwidth]{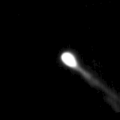}
    \includegraphics[width = 0.11\textwidth]{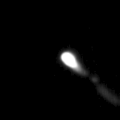}
    \caption{Experimental frames of Scan 39 in Case 01. Scan 39 is the last scan vector and only has four melt pool frames. The melt pool evolves from a relatively round shape to a tear shape.}
    \label{fig:MP_scan39_case01}
\end{figure}

In this section, we discuss the melt pool length evolution across the scan vectors. The average melt pool length of each scan vector is shown in Figure \ref{fig:scanwise_ave_length_CAPL}. From Scan 05 to Scan 39, melt pool lengths first increase, then plateau, and finally drop. This trend is consistent with the lengths of these scan vectors. The length of the scan vector starts as short with Scan 05 at the right bottom, reaches the maximum at Scan 19 and plateau, and decreases again to the last Scan 39. The scan vector averaged melt pool length is short when the total scan vector length is short and vice versa. This is because the melt pool needs the scan vector to be long enough to have its tail formed. On each scan vector, the melt pool always starts with a relatively round shape and gradually evolves into a tear shape as the laser moves (see Figure \ref{fig:MP_scan39_case01}). A scan vector that is too short will be dominated by the melt pool shape without tails.

The scan vector length is not the only factor that affects the melt pool length. In both the experiment data and modified CAPL simulation results, we observed some trends which could be explained by the influence of reheating from adjacent scan vectors. As shown in Scan 19-24 in Case 01, the difference in melt pool length between the neighbor scan vectors seems negligible. It seems the reheating from the adjacent scan vectors is negligible since Scan 24 has a comparable melt pool length to that in Scan 19. However, it is interesting to note that, even though they have identical scan vector lengths, Scan 39 always has a longer melt pool length compared to Scan 05. Note that Scan 05 is the first 45-degree scan vector and Scan 39 is likely due to the accumulation of residual heat in the scanning progress. Similar situations can be also found in other cases, see Figure \ref{fig:scanwise_ave_length_CAPL}. This also might be due to the reheating from the adjacent scan vectors, and this effect is only significant at the beginning of the scan vector where the melt pool is relatively round without a tail. 

\begin{table}
\centering

\footnotesize
\caption{Relative error of melt pool length predicted by CAPL}\label{table:err} 
\begin{tabular}{rrrrrrrrrrr}
 \hline
 Case No. &  01 &  02 &  03 &  04 &  05  \\
 Relative error (\%) & 11.11 & 9.80 & 9.62 & 8.81 & 8.44 \\
 \hline
 Case No. & 06 &  07 &  08 &  09 &  10 \\
 Relative error (\%) &8.21 & 8.87 & 9.03 & 11.66 & 12.70\\
\end{tabular}
\end{table}

\subsection{Melt pool width validation}
\begin{figure}[H]
    \centering
    \includegraphics[width=0.35\textwidth]{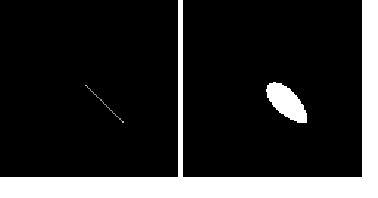}
    \caption{Melt pool from CAPL (left) and its interpolation (right).}
    \label{fig:IDW}
\end{figure}

\ch{The CAPL simulation results cannot be directly used to obtain the melt pool width because they are only defined on the scanning paths. In the present paper, we obtained the melt pool width from the reconstructed melt pool based on the CAPL simulation results.}
We reconstruct the melt pools by using inverse distance weighted interpolation of temperature distribution. Since the temperature for every element at any specific time is available, the temperature at any location can be given by interpolation:
\begin{align}
T(x) =  \frac{\sum_i^N  w_i(x)T_i}{\sum_i^N w_i(x)} 
\end{align}
where $w_i(x) = \frac{1}{d(x,x_i)^p}$ is the power $p$ of the inverse of distance between element $i$ and location $x$ (We use $p=1.3$ here \ch{which minimizes the errors}). \ch{$N$ is the number of elements inside the active body.} In the present paper, we interpolate the melt pool shape on a 0.96 mm $\times$ 0.96 mm region with a 120-pixel $\times$ 120-pixel frame. An example of such interpolation is given in Figure \ref{fig:IDW}. We obtain the melt pool width by the same approach used for experimental melt pool width with the interpolated temperature distribution: melt pool width is the length of the minor axis computed by the regionprops function with the given interpolated temperature distribution.

\begin{figure}
    \centering
    \includegraphics[width=0.48\textwidth]{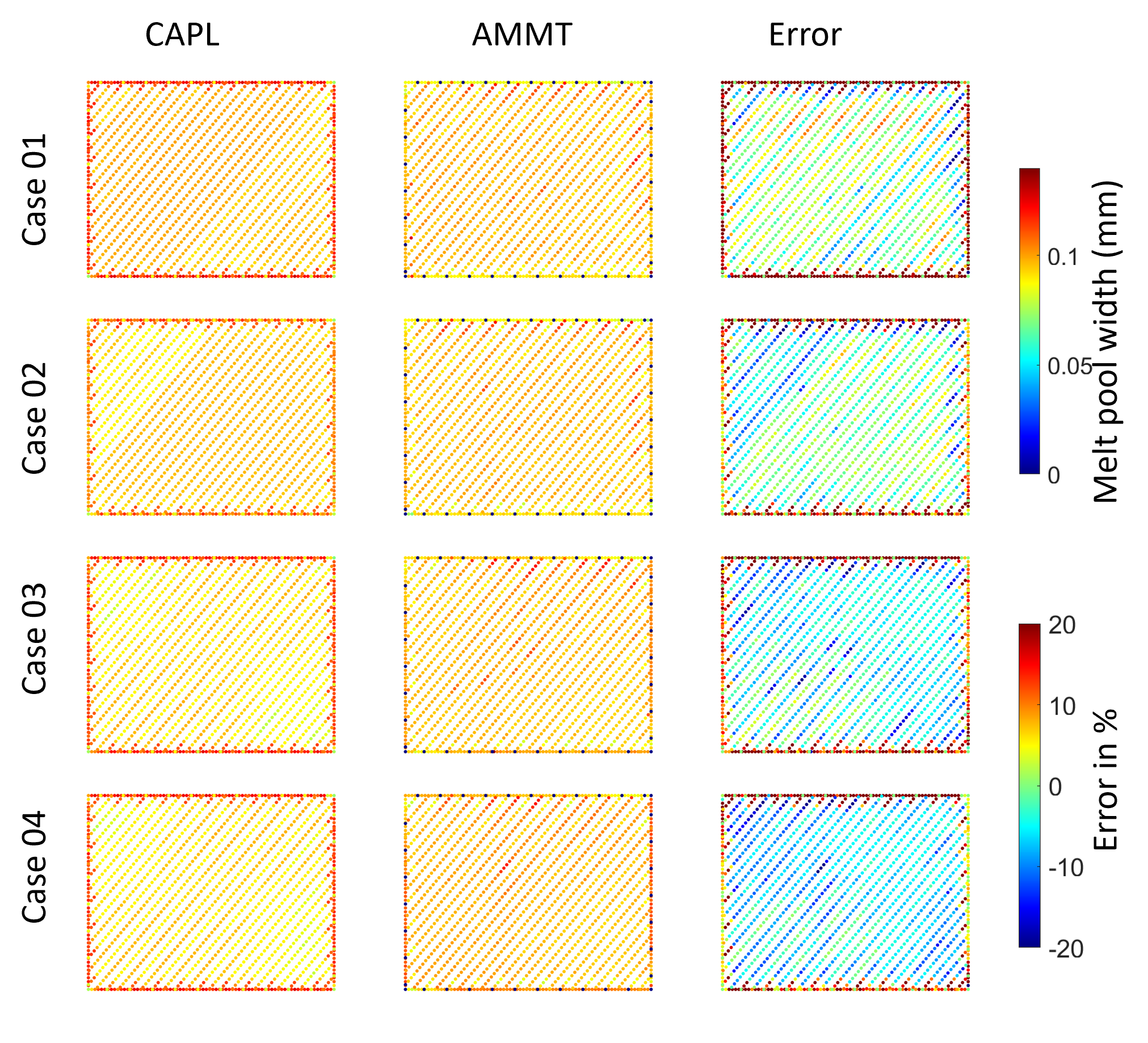}
    \caption{Experimental (AMMT) melt pool width map (middle), the interpolation results based on CAPL simulation results (left), and the relative errors in percentage (right) for cases 01 - 04. AMMT and CAPL results share the same color bar.}
    \label{fig:CAPL_widths}
\end{figure}

\begin{figure}
    \centering
    \includegraphics[width=0.4\textwidth]{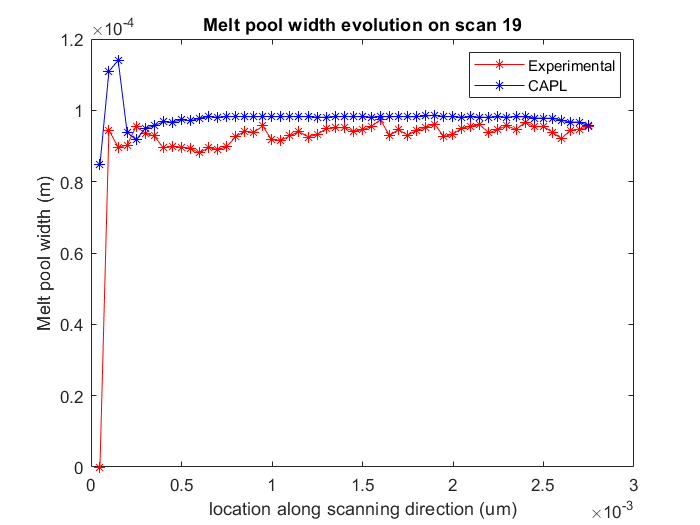}
    \includegraphics[width=0.4\textwidth]{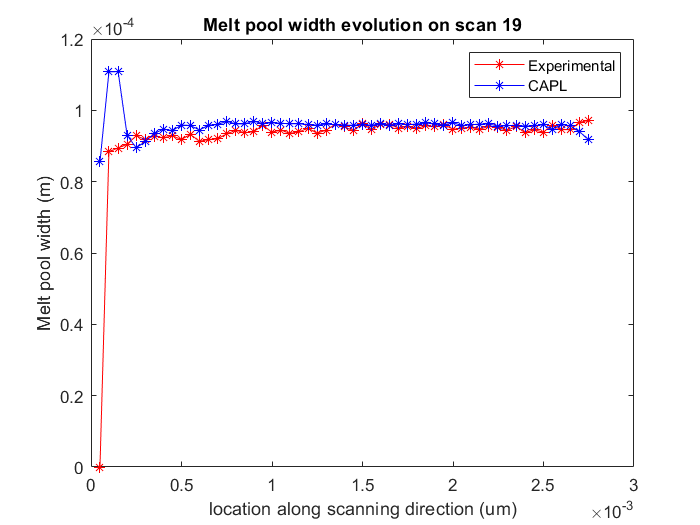}
    \includegraphics[width=0.4\textwidth]{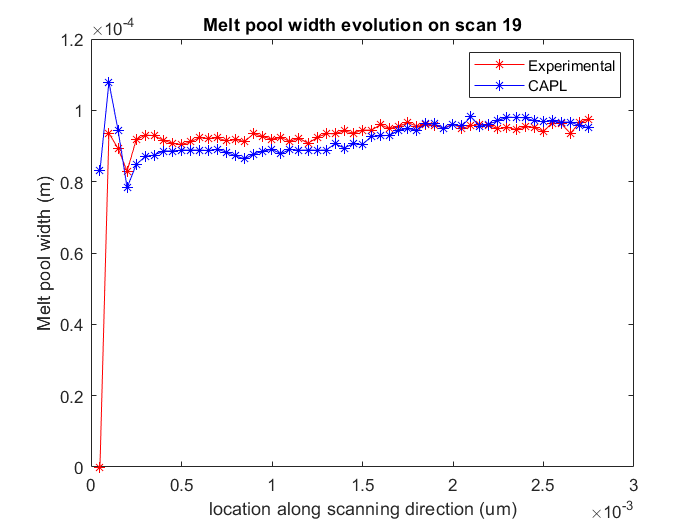}
    \includegraphics[width=0.4\textwidth]{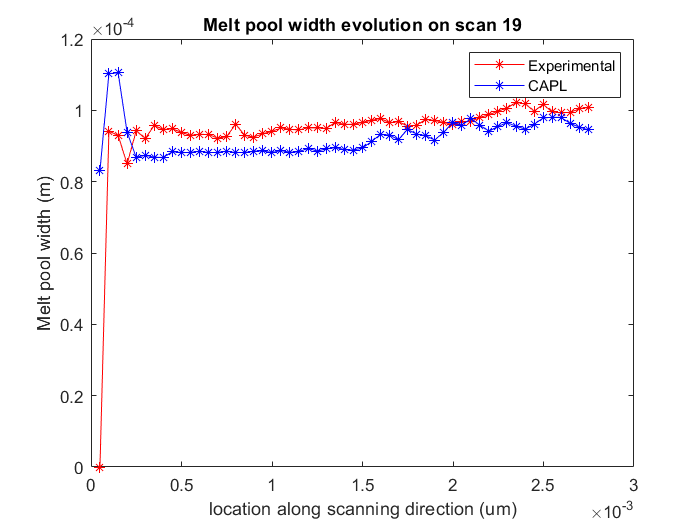}
    \caption{Melt pool width on Scan 19 of Cases 01, 02, 03, and 05. Both experimental data and CAPL simulation results show a relatively constant evolution.}
    \label{fig:width_scan19}
\end{figure}

Melt pool width from the experiment dataset and prediction by CAPL through interpolation and the relative errors are shown in figure \ref{fig:CAPL_widths}. The averaged relative errors for all cases are reported in table \ref{table:err_W}. The melt pool width has less variation compared with the melt pool length. For example, the melt pool width evolution of the Scan 19 is plotted as a function of scanning distance along the scan vector in Figure \ref{fig:width_scan19}). Note that laser power is not constant in Cases 02, 03, and 05, but both CAPL simulation results and experimental data suggest melt pool width has very little variation compared with the melt pool length in a scan vector like it is in Case 01 (see Figure \ref{fig:scan19_24_case01}), \ch{indicating that the melt pool width is less sensitive to laser power than the melt pool length}.

\begin{table}
\centering

\footnotesize
\caption{Relative error of melt pool width predicted by interpolation of CAPL results}\label{table:err_W} 
\begin{tabular}{rrrrrrrrrrr}
 \hline
 Case No. &  01 &  02 &  03 &  04 &  05 \\
 Relative error (\%) & 9.80	& 6.93	& 8.62	& 8.753	& 9.50	\\
\hline
 Case No. & 06 &  07 &  08 &  09 &  10 \\
 Relative error (\%)	& 9.90	& 10.64	& 9.99& 	13.71& 	9.36\\
\end{tabular}
\end{table}

\section{Conclusion and future directions}
\label{sec:conclusion}

\q{We developed an improved path-scale PBF thermal simulation approach based on contact-aware path-level (CAPL) discretization.} We validate the proposed approach with melt pool shapes acquired by the Additive Manufacturing Metrology Testbed (AMMT) built by the National Institute of Standards and Technology (NIST). We demonstrated that the proposed approach achieves a good match of the melt pool length and width compared with the experimental data. We also discussed the influence of laser power on the thermal history at the path-scale level.

\begin{figure}
    \centering
    \includegraphics[width = 0.4\textwidth]{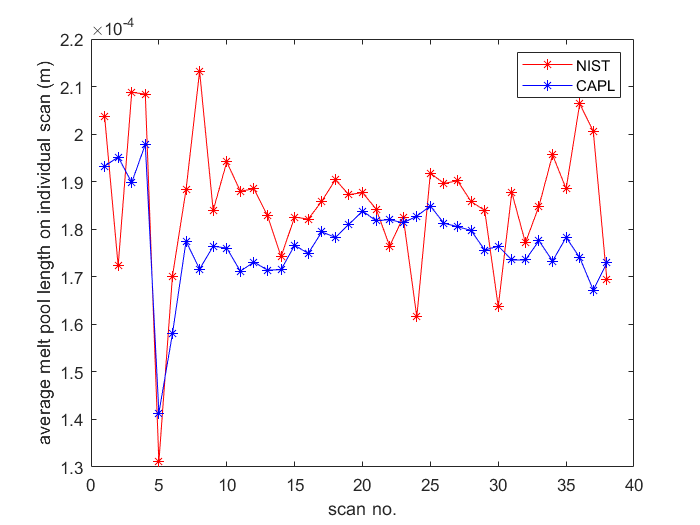}
    \caption{Scan-wise average melt pool length, Case 09. Relatively large errors are observed at around Scan 05 and Scan 39, where high laser power is used.}
    \label{fig:scanwise_ave_length_CAPL_case09_10}
\end{figure}

\ch{One source of error during the CAPL validation is the increase in the difference in melt pool length at the beginning of the laser scan while the laser power and speed remain constant in Case 01. This is likely due to the use of constant absorptivity in the current work, which is widely assumed among LPBF thermal history simulations \cite{zhang2019resolution, mayi2021transient, huang2021efficient}. Such an assumption results in an underestimation of the absorptivity at the beginning of high-power scan vectors during the calibration, for example in Case 01 and Case 09 (see Figure \ref{fig:scanwise_ave_length_CAPL_case09_10}). } Experimental measurements have observed absorptivity to change throughout the laser scan path \cite{trapp2017situ,lane2020transient,simonds2018time}. For example, a distinctive four-stage change in laser absorptivity has been identified for heating a stainless steel plate \cite{simonds2018time}: (a) an initial rise due to the shiny plate melting increasing surface roughness, (b) a subsequent drop due to the liquid melt pool formation and increased reflectivity, (c) the second rise due to the formation of the keyhole, and (d) eventually a keyhole-related high-frequency periodic oscillation. \ch{A dynamic laser absorptivity depending on the time and distance of the laser start as well as the condition inside the meltpool therefore will better capture the melt pool behavior near the beginning of the laser scan.}

\begin{figure}
    \centering
    \includegraphics[width = 0.47\textwidth]{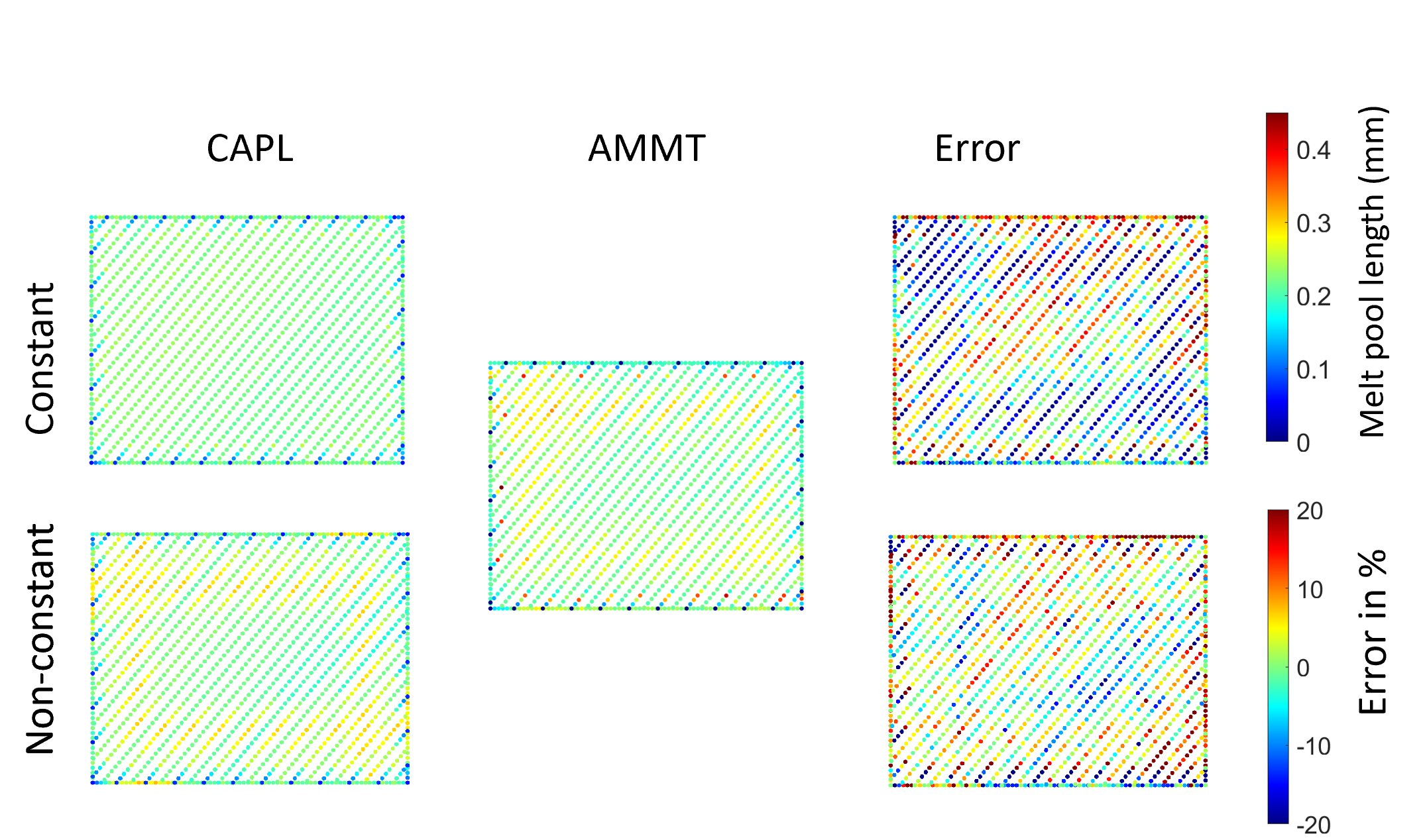}
    \caption{CAPL results of Case 01 with the non-constant absorptivity surrogate model (higher absorptivity at beginning of every scan) on the bottom and the results with constant absorptivity on top. Mean error is smaller (9.10\%) compared with no surrogate model (11.11\%) shown in table \ref{table:err}. The underestimation of the "bump" can be seen improved with the non-constant model.}
    \label{fig:mod_a_length_map}
\end{figure}

\begin{figure}
    \centering
    \includegraphics[width = 0.4\textwidth]{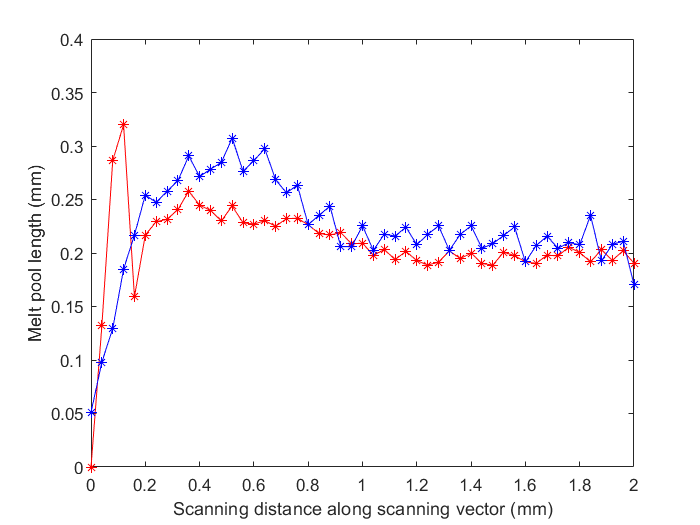}
    \includegraphics[width = 0.4\textwidth]{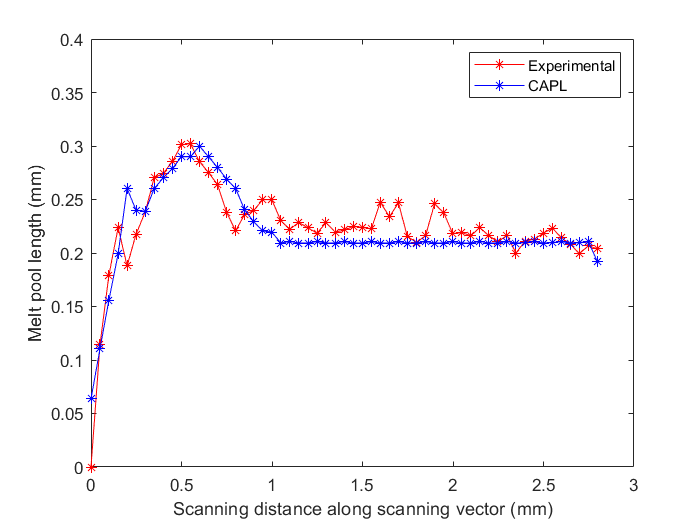}
    \includegraphics[width = 0.4\textwidth]{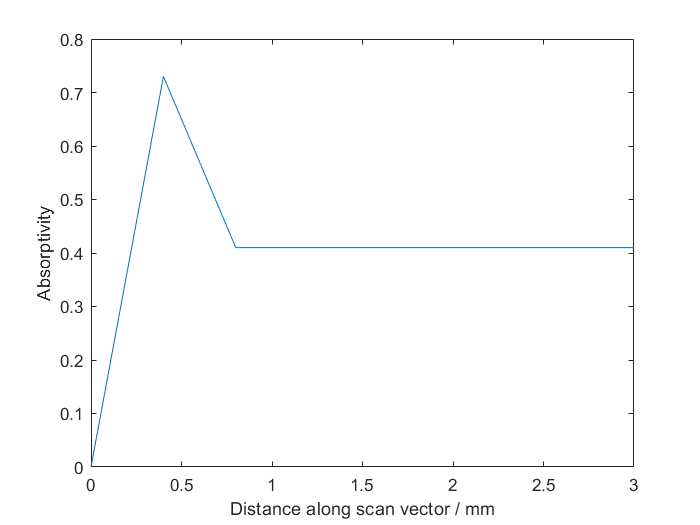}
    \caption{CAPL results of Case 01 with the surrogate model of Scan 01 (top) and Scan 19 (middle). The ``bump'' can be reproduced by the piecewise linear surrogate absorptivity model (bottom). The peak value of the surrogate model is 0.73.}
    \label{fig:mod_a_length}
\end{figure}

Based on this observation, we create a simplified model that captures the laser absorptivity trend in stages (a) and (b). We assign the absorptivity by a piecewise linear model (see Figure \ref{fig:mod_a_length}) in the range from 0 to 0.8 mm. A constant absorptivity (0.41) is assigned to the rest of the scan vector. We apply this simplified model to Case 01. The results are shown in Figure \ref{fig:mod_a_length_map}. The dynamic laser absorptivity model not only results in a better overall prediction of the melt pool length but also reproduces the ``bump'' seen at the beginning of the scan (see Figure \ref{fig:mod_a_length}). A more accurate laser absorptivity that is based on physical testings and models the keyhole behaviors in stages (c) and (d) is outside the scope of the current discussion and will be studied as part of future work to further improve the CAPL approach.



\begin{figure}
    \centering
    \includegraphics[width = 0.47\textwidth]{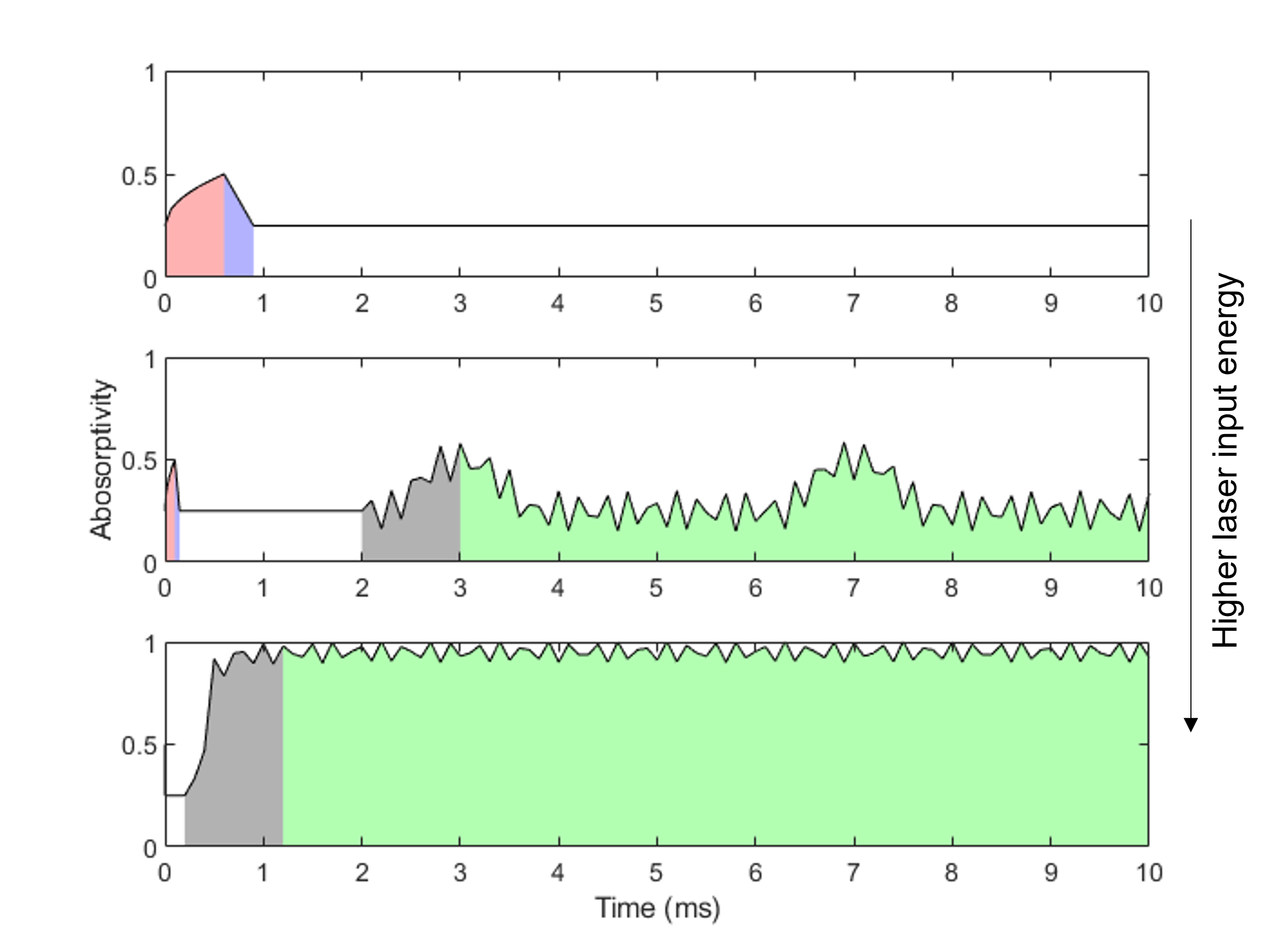}
    \caption{A schematic of dynamic laser absorptivity reproduced from paper \cite{simonds2018time}. The top, middle, and bottom are the dynamic absorptivity under the conduction model, transition model, and keyhole model. The different models are determined by the laser input energy density. Stages (a, in red) and (b, in blue) are mostly visible in the conduction model, while in keyhole model is dominated by stages (c, in black) and (d, in green). In the transition model, all four stages are visible and there is low-frequency keyhole oscillation.}
    \label{fig:ab_model}
\end{figure}

\begin{figure}
    \centering
    \includegraphics[width = 0.45\textwidth]{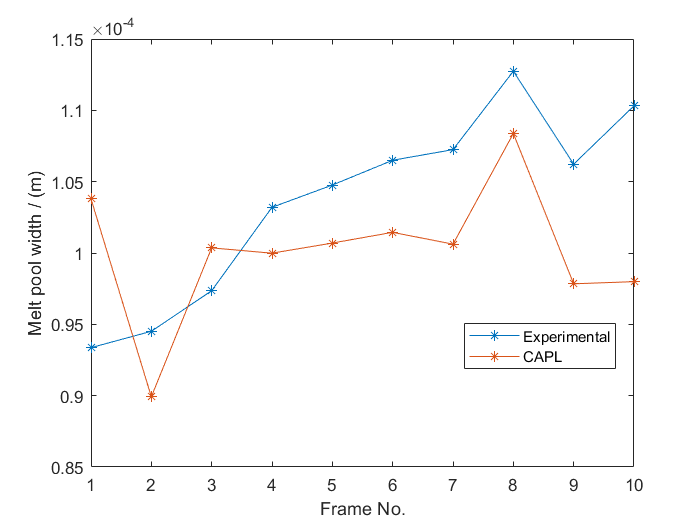}
    \caption{Average of melt pool width of Scan 01, Case 01-Case 10.}
    \label{fig:ave_width_scan01}
\end{figure}

\begin{figure}
    \centering
    \includegraphics[width = 0.4\textwidth]{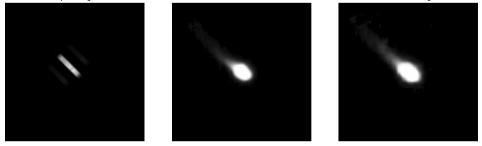}
    \caption{An example of melt pool prediction by machine learning. The left, middle, and right are the input, ground truth, and prediction respectively.}
    \label{fig:cgan}
\end{figure}

The second future improvement is \ch{to better predict} the melt pool width as well as the shape of the melt pool. Recent progress in machine learning has enabled an exponential increase in research into leveraging machine learning to predict and control additive manufacturing processes \cite{wang2020machine,razvi2019review,regenwetter2022deep}. We are exploring a conditional Generative Adversarial Network (cGAN) -based approach \cite{goodfellow2020generative, isola2017image, chen2021geometry} to predict the melt pool image based on the thermal history along the laser scan path. GAN uses its generator to generate candidates and uses its discriminator to evaluate the candidates. The generator and the discriminator are trained so that eventually the generator is able to predict accurate melt pool frames. The data pair of thermal history and melt pool images for neural network training and validation are obtained through CAPL and AMMT data, respectively (see Figure \ref{fig:cgan}). Specifically, we snapshot the thermal history into a series of thermal distribution images that matches the field-of-view and time steps of the captured melt pool images to work as the conditional inputs into the cGAN. Compared to the interpolation approach discussed in Section \ref{sec:validation}, the machine learning-based approach better captures the overall shape of the melt pool, which, in turn, provides further insight into the steep thermal gradient inside the melt pool. We will validate the prediction of the melt pool images through geometric characteristics, including length, width, and angle, of the melt pool as part of future work. \ch{Such a machine learning model could be used to improve the modified conduction model in the present paper. For example, here the conduction characteristic distance $d_0$ is a constant, and its machine learning model could be used to map this quantity as a function of process parameters to further improve accuracy.}

%

\section*{Acknowledgements}
The authors thank Ho Yeung from the National Institute of Standards and Technology for providing the melt pool monitoring data and helpful discussions. This research was supported by the National Institute of Standards and Technology. The responsibility for errors and omissions lies solely with the author. 

\section*{References}

\bibliography{ref}

\end{document}